\documentclass[final,3p,times,twocolumn]{elsarticle}
\usepackage{graphicx}
\usepackage{amssymb}
\usepackage{amsmath} 

\usepackage{lipsum}
\makeatletter
\def\ps@pprintTitle{%
 \let\@oddhead\@empty
 \let\@evenhead\@empty
 \def\@oddfoot{}%
 \let\@evenfoot\@oddfoot}
\makeatother

\begin{document}
\begin{frontmatter}
\title{Measuring ligand-receptor binding kinetics and dynamics using k-space image correlation spectroscopy}
\author[label1]{Hugo B. Brand\~ao}
\author[label1]{Hussain Sangji}
\author[label1]{Elvis Pand\v{z}i\'c}
\author[label3]{Susanne Bechstedt}
\author[label3]{Gary J. Brouhard}
\author[label1,label2]{Paul W. Wiseman}
\address[label1]{Department of Physics, McGill University, Montr\'eal, Qu\'ebec, H3A 2T8, Canada}
\address[label2]{Department of Chemistry, McGill University, Montr\'eal, Qu\'ebec, H3A 2K6, Canada}
\address[label3]{Department of Biology, McGill University, Montr\'eal, Qu\'ebec, H3A 1B1, Canada}
\begin{abstract}
Accurate measurements of kinetic rate constants for interacting biomolecules is crucial for understanding the mechanisms underlying intracellular signalling pathways. The magnitude of binding rates plays a very important molecular regulatory role which can lead to very different cellular physiological responses under different conditions. Here, we extend the k-space image correlation spectroscopy (kICS) technique to study the kinetic binding rates of systems wherein: (a) fluorescently labelled, free ligands in solution interact with unlabelled, diffusing receptors in the plasma membrane and (b) systems where labelled, diffusing receptors are allowed to bind/unbind and interconvert between two different diffusing states on the plasma membrane. We develop the necessary mathematical framework for the kICS analysis and demonstrate how to extract the relevant kinetic binding parameters of the underlying molecular system from fluorescence video-microscopy image time-series. Finally, by examining real data for two 
model experimental systems, we demonstrate how kICS can be a powerful tool to measure molecular transport coefficients and binding kinetics.
\end{abstract}
\end{frontmatter}
\section{Introduction}
\label{intro}

\begin{figure}[!ht]
\centering
\includegraphics[scale=0.23]{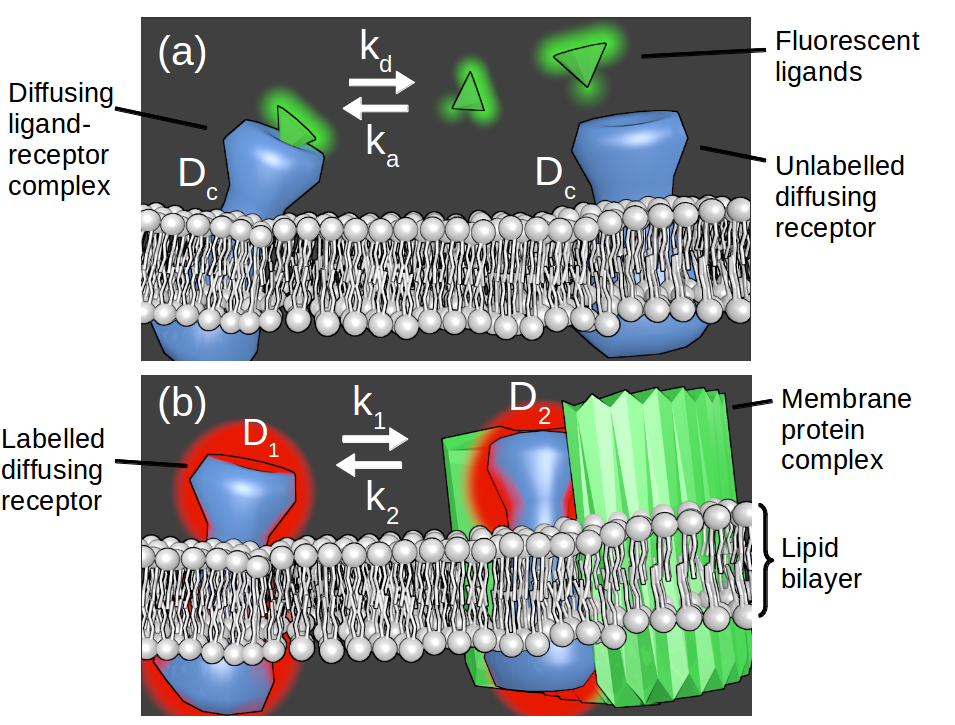}
\caption{Schematic illustration of two kinetic models: (a) Fluorescently tagged, freely diffusing ligands in solution bind and unbind to unlabelled membrane receptors with kinetic rates $k_a$ and $k_d$. Unbound ligands diffuse quickly and only contribute to a fluorescence background signal. Ligand-receptor complexes diffuse slowly with coefficient $D_c$ and and are visible in the imaging plane (b) Fluorescently tagged, diffusing receptors with diffusion coefficient $D_1$ bind and unbind to other membrane receptors to form complexes with kinetic rates $k_1$ and $k_2$. Protein complexes diffuse with a different coefficient, $D_2$.}
\label{fig::kinetic_mechanisms}
\end{figure}

Virtually all aspects of the cellular life-cycle are governed by signalling cascades. Early signalling events occuring at the plasma membrane are of particular interest; the rates of ligand-receptor binding and receptor-receptor binding have been shown to play a key role in the activation of the immune response, cell growth, motility, and death~\cite{Feinerman2008} \cite{Govern2010} \cite{Dushek2009} \cite{Kholodenko1999} \cite{Curtis2001} \cite{Kaufmann2012} \cite{Elmore2007}. It is clear that physiological cell response is intricately tied via feed-back mechanisms to the biochemical networks which make up the cell. As a result, we must have good experimental techniques to probe the dynamics and kinetics of biomolecular interactions in situ in living cells to fully understand cell function. 

Fluorescence based methods have been some of the most commonly used tools to probe in-vivo intracellular signalling. Common techniques such as fluorescence recovery after photobleaching (FRAP) \cite{MonteroLlopis2012}, F\"orster resonance energy transfer (FRET), fluorescence lifetime imaging (FLIM)~\cite{Yasuda2006}, and fluorescence correlation spectroscopy (FCS) each make it possible to study the interactions of biomolecules in-situ \cite{Michelman-Ribeiro2009} \cite{Stasevich2010}.

With FCS, a laser beam focus defines a small observation volume in a cell and FCS correlates fluorescence intensity fluctuations as labelled molecules move in and out of focus. Over the years, this method developed by Elson, Magde, and Webb~\cite{Elson1974} \cite{Magde1972} has been extended to accommodate the development of fluorescence imaging technology such as confocal and total-internal-reflection-fluorescence (TIRF) microscopy. 

There is a technique of particular interest for the study of surface reaction kinetics called total-internal-reflection fluorecence correlation spectroscopy (TIR-FCS).  The working principle is similar to the original FCS experiments, whereby fluctuations in fluorescence intensity are collected, and the time-series is used to calculate correlation functions. The correlation functions are subsequently used to infer chemical reaction rates for the reversible association of fluorescently tagged ligands as they adhere to stationary surface binding sites on the sample substrate~\cite{Thompson1981}\cite{Thompson1983}\cite{Starr2001}. 

Unlike conventional FCS experiments, where fluctuations are measured from within a 3D, diffraction-limited excitation volume (often assumed Gaussian in shape), TIR-FCS measures fluctuations within an excitation plane. In the lateral (imaging) plane, which is parallel to the the sample substrate, the fluorescence illumination is relatively uniform. In the axial direction (perpendicular to the plane), the luminance intensity falls off approximately exponentially which is a characteristic property of totally-internally reflected light called evanescence~\cite{Thompson2011}. Typically, the fluorescence excitation depth extends only about 100~nm into the sample~\cite{Thompson2011}; this creates a fluorescence imaging modality where background fluorescence is strongly attenuated which allows for better resolution of features near the sample substrate.

A closely related set of techniques to FCS is known as image correlation spectroscopy (ICS). The ICS techniques represent the extension of FCS into the spatial domain. Spatio-temporal image correlation spectroscopy (STICS) and k-space image-correlation spectroscopy (kICS) are examples of two ICS techniques which simultaneously take into account both spatial and temporal correlation information. By measuring fluctuations in pixel intensities of an image time-series, STICS and kICS can measure the transport dynamics and photophysical kinetics of fluorescently tagged molecules~\cite{Kolin2007}. It should be noted that TIR-FCS was also extended into the spatial domain and has also been shown to recover transport dynamics similarly to ICS~\cite{Kannan2007}~\cite{Sankaran2009}. 

STICS is often used to measure the flow of proteins within cells in order to create vector maps of molecular transport, but the technique can also be used to measure diffusion coefficients and number densities of fluorescent molecules~\cite{Hebert2005}. Recent developments of STICS now permit analysis of two-colour imaging data to measure co-localization and co-transport of molecular species~\cite{Toplak2012}. Tanaka and Papoian have also recently suggested a method to exploit the spatio-temporal information embedded in the STICS autocorreltion function in order to measure the binding kinetic paramters of reaction-diffusion systems~\cite{Tanaka2010}. The authors use an iterative computational simulation coupled to a minimization scheme approach in order to identify appropriate reaction-diffusion models and quantify the model parameters. 

The two spatio-temporal image correlation techniques (STICS and kICS) are very similar in nature, but there are some small key differences. The working principle of kICS is nearly identical to STICS, and much of the same information can be measured; however, kICS does not directly correlate fluctuations in image pixel intensities, but instead, it calculates time-correlation functions from spatially Fourier-transformed image series. 

There are distinct advantages of operating in Fourier space (k-space). Firstly, as shown by Kolin et al.~\cite{Kolin2006}, kICS can recover phototophysical kinetics and transport dynamics independently of the shape of the optical point-spread-function (PSF). Secondly, kICS can readily separate the purely temporal kinetics from the spatio-temporal dynamics without requiring non-linear fitting for simple, one-component systems. As we will show here, the second property will allow us to accurately measure chemical kinetic association rates of non-interacting one-component model systems where fluorescence fluctuations arise from binding in and out of the focal plane (like in TIR-FCS). 

The motivation of this work is to outline a new method to measure chemical reaction kinetics using TIRF microscopy images as input. We present the basic theory and show how even in the presence of complicated fluorescence photophysics, such as probe blinking and photobleaching, it may be possible to extract kinetic reaction parameters. The original work done by Kolin et al. derived the functional form of the kICS correlation function for multi-component, non-interacting chemical species~\cite{Kolin2006}. Here, we further extend this work to allow for the possibility of two interacting populations. It is advantageous to use kICS for two-component interacting systems, since the solutions to the coupled differential equations take a simpler form in k-space. STICS, and the spatial extension of TIR-FCS, will require more elaborate fitting routines since the autocorrelation functions will take on a more complex form.

We model two kinds of chemical kinetic mechanisms as illustrated in Fig.~\ref{fig::kinetic_mechanisms}. The first system (Fig.~\ref{fig::kinetic_mechanisms}~a) is that of binding kinetics of a one-component system occuring in and out of the imaging plane. A typical biological scenario would be that of freely diffusing ligands in the cytoplasm (or in the extra-cellular matrix) binding to receptors diffusing in the membrane.  
The second system (Fig.~\ref{fig::kinetic_mechanisms}~b) is that of binding kinetics occuring within the imaging plane. In this reaction-diffusion system, a molecule is confined to exist in one of two diffusing states and may dynamically switch between them. This can represent the docking of receptors with other receptors or the docking of receptors to cytoskeletal elements near the plasma membrane.

For both kinetic model systems, we will show how to perform the image-processing steps involved in extracting the kinetic binding and unbinding rates from these equilibrium fluctuations.

\section{Theory and Methods}
\begin{figure*}[!ht]
\centering
\includegraphics[scale=0.29]{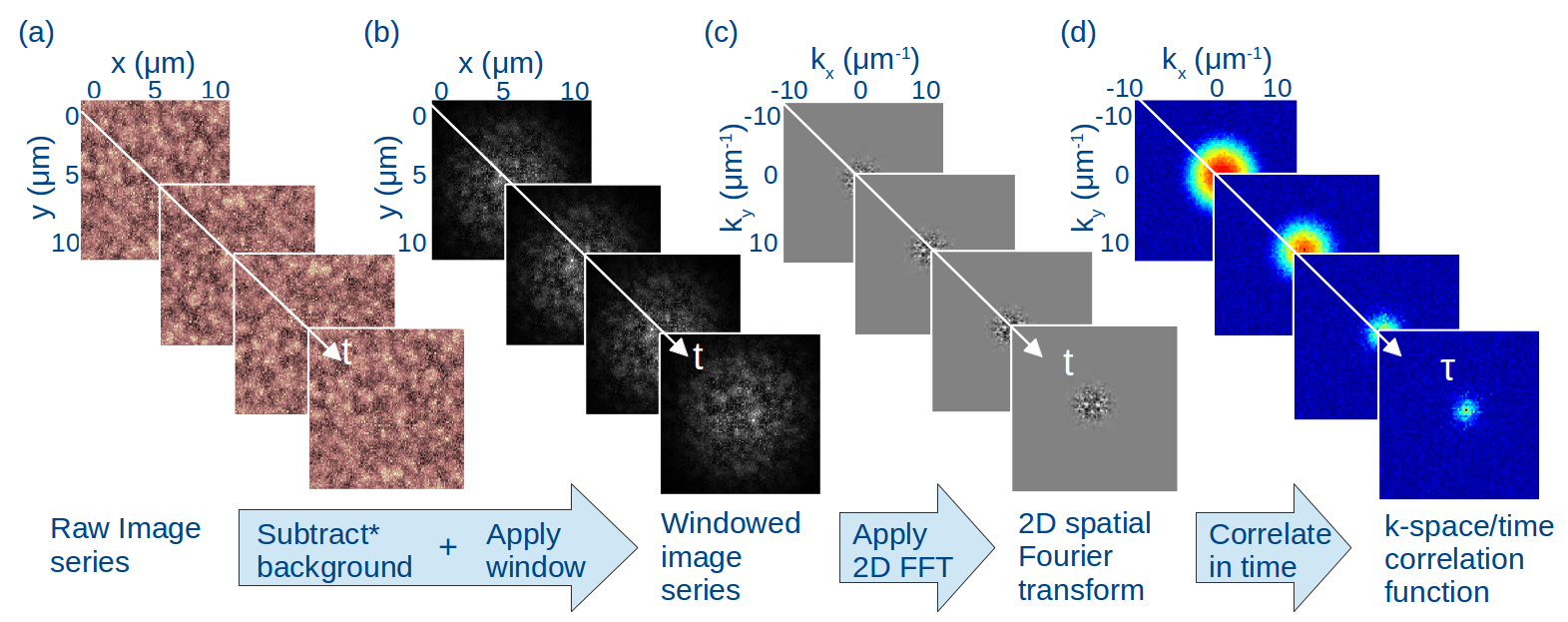}
\caption{The image processing steps required to obtain the k-space image correlation function. (a) An image time-series is obtained from TIRF video-microscopy of diffusing fluorescent species in a focal plane. (b) An optional subtraction of the mean image (averaged in time) can help to remove immobile fluorescent species before a window function is applied to each image. (c) Each image in the time-series is Fourier spatially transformed. The spatial window function is used to minimize the effects of spectral leakage prior to this step. (d) Fourier transformed images are correlated in time and normalized by the zeroth time-lag (the first frame of the correlated image series).}
\label{fig::kicsMethod}
\end{figure*}

\subsection{kICS: Theory}
In ICS we are fundamentally interested in how the apparent concentration of a fluorescent species changes in time at a particular point in space; these fluctuations can arise from biochemical reactions or simply movement of the chemical species in and out of the focal volume. In k-space, ICS simply measures how spatial frequency intensities change (or fluctuate) in time, where images in the microscopy series have been spatially Fourier transformed. For a freely-diffusing, non-interacting particle in 2D, the k-space/time image correlation function, $r({\bf{k}};\tau,t)$, has the form 
\begin{equation}
\label{eq::icf}
r({\bf{k}};\tau) = Nq^2 ~|\tilde{I}({\bf{k}})|^2 ~\langle\theta(0)\theta(\tau) \rangle ~ e^{i{\bf{k}\cdot v\tau - |k|}^2 D\tau}
\end{equation} 
\noindent for $|{\bf k}| > 0$, where ${\bf k}$ is the spatial wave-vector, $N$ is the average number of particles in an image, $q$ is the fluorescent probe's quantum yield, ${\bf v}$  is the flow velocity of the imaged molecular species, and $D$ is the diffusion coefficient. $\tilde{I}({\bf{k}})$ is the 2D Fourier transform of the imaging system's point-spread function, and $\langle\theta(0)\theta(\tau) \rangle $ is the photophysics temporal correlation function; we assume these depend solely on space and time respectively.

Eq.~\ref{eq::icf} is experimentally computed by spatially Fourier transforming a time-series of fluorescence microscopy images before correlating the Fourier images in time as illustrated in Fig.~\ref{fig::kicsMethod}. Because in practice we have finite-sized image series, we typically apply a windowing function to the images before taking their spatial Fourier transform. The window function (such as a Hann or Hamming window) is used to minimize the effect of spectral leakage, which can introduce some bias in the quantities computed from the correlation function~\cite{Hamming}. 
The great advantage of kICS is that in Fourier space we can easily separate variables depending on space-only, on time-only, and on space-time. Because the contribution to the correlation function of the optical PSF is only space dependent, we can remove it by normalization; we divide $r({\bf k};\tau)$ by the zero-time-lag correlation function, $r({\bf k};0)$. This leaves us with:
\begin{equation}
\label{eq::icf_norm}
\frac{r({\bf{k}};\tau)}{r({\bf{k}};0)} = G(\tau)~ e^{i{\bf{k}\cdot v\tau - | k |}^2 D\tau}
\end{equation} 
\noindent where $G(\tau) = \langle\theta(0)\theta(\tau) \rangle/\langle\theta(0)^2\rangle$ is the normalized photophysics correlation function. The result is that the dependence on $\tilde{I}({\bf{k}})$, $q$, and $N$ drops out meaning we can measure binding kinetics and transport dynamics without first calibrating the system to find out the shape of the point spread-function~\cite{Kolin2006}. In the sections that follow, we demonstrate how to use Eq.~\ref{eq::icf_norm} to measure kinetic parameters arising from molecular interactions.

\subsection{Ligand-receptor binding}
The main photophysical contributions for the model dynamic system of Fig.~\ref{fig::kinetic_mechanisms}~a) are captured by considering three main sources of fluorescence intensity fluctuations: 1) a fluorescently tagged ligand in solution binds to its receptor on the cell membrane, 2) the fluorescent label blinks due to excitation and depletion of excited electronic states, and 3) the fluorescent marker photobleaches irreversibly. Labelling each of these respective photophysics contributions as $\psi_{bind}(t)$, $\phi_{blink}(t)$, and $\phi_{photo}(t)$, the fluorescence signal as a function of time, $\theta(t)$, of a single fluorescently labelled molecule is given by the product of the three time-dependent signals. That is,
\begin{equation}
\theta(t) = \psi_{bind}(t) \phi_{blink}(t) \phi_{photo}(t)
\end{equation} 
Each of these photophysical variables can take a value of $1$, if the molecule is emitting/visible, or $0$ otherwise. The change occurs stochastically given by some probability distribution function which we may assume is a memoryless (Markovian) process. The measured time-correlation function, $G(\tau)$, (for the stationary random process) is then given by: 
\begin{equation} \label{eq::gtau} 
G(\tau) = \langle \psi_{bind}(0)\psi_{bind}(\tau)\rangle ~\cdot ~ \Gamma_{fluor}(\tau) 
\end{equation}
The angular brackets $\langle ... \rangle$ denotes a time-average (or ensemble average) and we have assumed here that the fluorescence photophysics is independent of the binding kinetics. The fluorescence photophysics correlation function is:
\begin{equation} \label{eq::gtau_fluorescence}
\Gamma_{fluor}(\tau) = \langle  \phi_{blink}(0)\phi_{blink}(\tau) \phi_{photo}(0)\phi_{photo}(\tau) \rangle 
\end{equation}
When control measurements of the fluorescence photophysics can be used to obtain $\Gamma_{fluor}(\tau)$ separately from the binding kinetics (or when fluorescence photophysics is negligible, i.e. $\Gamma_{fluor}(\tau)\approx 1$), it is possible to measure binding kinetic rate parameters from $G(\tau)$. The chemical kinetic rates are extracted from $\langle \psi_{bind}(0)\psi_{bind}(\tau)\rangle$, which is measured from Eq.~\ref{eq::icf_norm} as we let $|{\bf k}| \rightarrow 0 $ (this is typically done by extrapolation to the zero k-point). 
For the two-state binding model system of Fig~\ref{fig::kinetic_mechanisms}~a),
\begin{equation} \label{eq::gtau_binding}
\langle \psi_{bind}(0)\psi_{bind}(\tau)\rangle = \frac{k_a}{k_a+k_d} + \frac{k_d}{k_a+k_d} e^{-(k_a+k_d)\tau}
\end{equation}
\noindent where $k_a$ and $k_d$ are the association and dissociation rate constants we wish to measure.

The implicit assumption made here is that freely diffusing ligands are not visible unless they are bound to their respective receptors. For TIRF video-microscopy this is a valid assumption since we may assume that unbound ligands, due to their high diffusion coefficients, diffuse quickly out of the small focal volume in the direction perpendicular to the imaging plane. Due to the slow time-response of the detector, they appear as a blur in solution and contribute only to a diffuse background signal. When they bind to a slowly moving receptor, the ligand-receptor complex becomes resolvable and visible as a point-source of light. We further assume the ligand concentration is approximately constant; in this scenario, $k_a$ will depend linearly on the ligand concentration~\cite{Lauffenburger} (See ~\ref{sec::One_component_appendix} for details). 

In this treatment, we have assumed that correlations arising from the lateral diffusion of ligand-receptor complexes out of the image region-of-interest are small and do not contribute to $G(\tau)$. This approximation is valid for images with large regions-of-interest and/or slow receptor diffusion compared to the kinetic binding rates. Also, in the case where ligand-receptor complexes can diffuse in the membrane but are confined laterally to remain in the imaging region-of-interest, the approximation becomes exact.

\subsection{Receptor-receptor docking}

For the model system of Fig.~\ref{fig::kinetic_mechanisms}~b), we are observing a reaction-diffusion system wherein a fluorescently tagged protein on the cell membrane is interconverting between two diffusing states. The fast state has diffusion coefficient $D_1$ and the slow state has diffusion coefficient $D_2$. The two-state diffusion system stochastically interconverts from state 1 to 2 with the forward kinetic rate $k_1$ and reverse rate $k_2$ as illustrated in the figure. This kind of system could occur following the docking of the protein with another kind of receptor. The assumption made in this case is that there is an abundance of the second kind of receptor such that the concentration of free receptors is approximately constant. The reason for this assumption is that the effective kinetic rates $k_1$ and $k_2$ will depend on the diffusion coefficient of the protein itself as well as the free receptor concentration as shown previously by Lauffenburger and Linderman~\cite{Lauffenburger}.

The derivation of concentration fluctuations of two interacting diffusing species is solved by Berne and Pecora~\cite{BernePecora}. We accomodate their treamtent to include fluorescence photophysics. Because the full solution of the coupled differential equations takes on a very complex form, we consider only some limiting cases which can be used to extract our parameters of interest (for a more detailed treatment, see ~\ref{sec::Two_component_appendix}).

In one scenario we look only at very small values of $|{\bf{k}}|^2$ (i.e. correlations on large spatial scales); this is the so-called fast-exchange regime. Here we effectively measure only averaged molecular diffusion as the reaction time is much shorter than the characteristic diffusion time at these length scales . Mathematically, this is expressed as the regime where $(k_1 + k_2) >> D_1 |{\bf k}|^2$. At small temporal lags, $\tau$, and small $|{\bf{k}}|^2$, the zeroth time-lag normalized kICS correlation function then takes the form:
\begin{equation} \label{eq::fast_exchange}
 r_{fast}(|{\bf{k}}|^2,\tau) = C_1 ~ G(\tau)  \exp(-|{\bf k}|^2D_{\mathrm{eff}}\tau)
\end{equation}
\noindent where, 
\begin{equation}
 D_{\mathrm{eff}}=\frac{k_2D_1 +k_1D_2}{k_1 + k_2}
\end{equation}
\noindent is a measured effective diffusion coefficient and $C_1$ is just a constant. 
In the limit where we look at large values of $|{\bf{k}}|^2$ (i.e. correlations on small spatial scales), some of the finer details of the molecular interactions start to emerge. This is the so-called slow-exchange regime, which occurs when $(k_1 + k_2) << D_1 |{\bf k}|^2$.  At large temporal lags, when $\tau >> (D_1 |{\bf k}|^2)^{-1}$, the normalized kICS correlation function takes the form:
\begin{equation} \label{eq::slow_exchange}
 r_{slow}(|{\bf{k}}|^2,\tau) = C_2 ~G(\tau)~\exp(-|{\bf{k}}|^2 D_2\tau-k_2\tau)
\end{equation}

\noindent where, $C_2$ is a constant.

\begin{figure*}[!ht]
\centering
\includegraphics[scale=0.35]{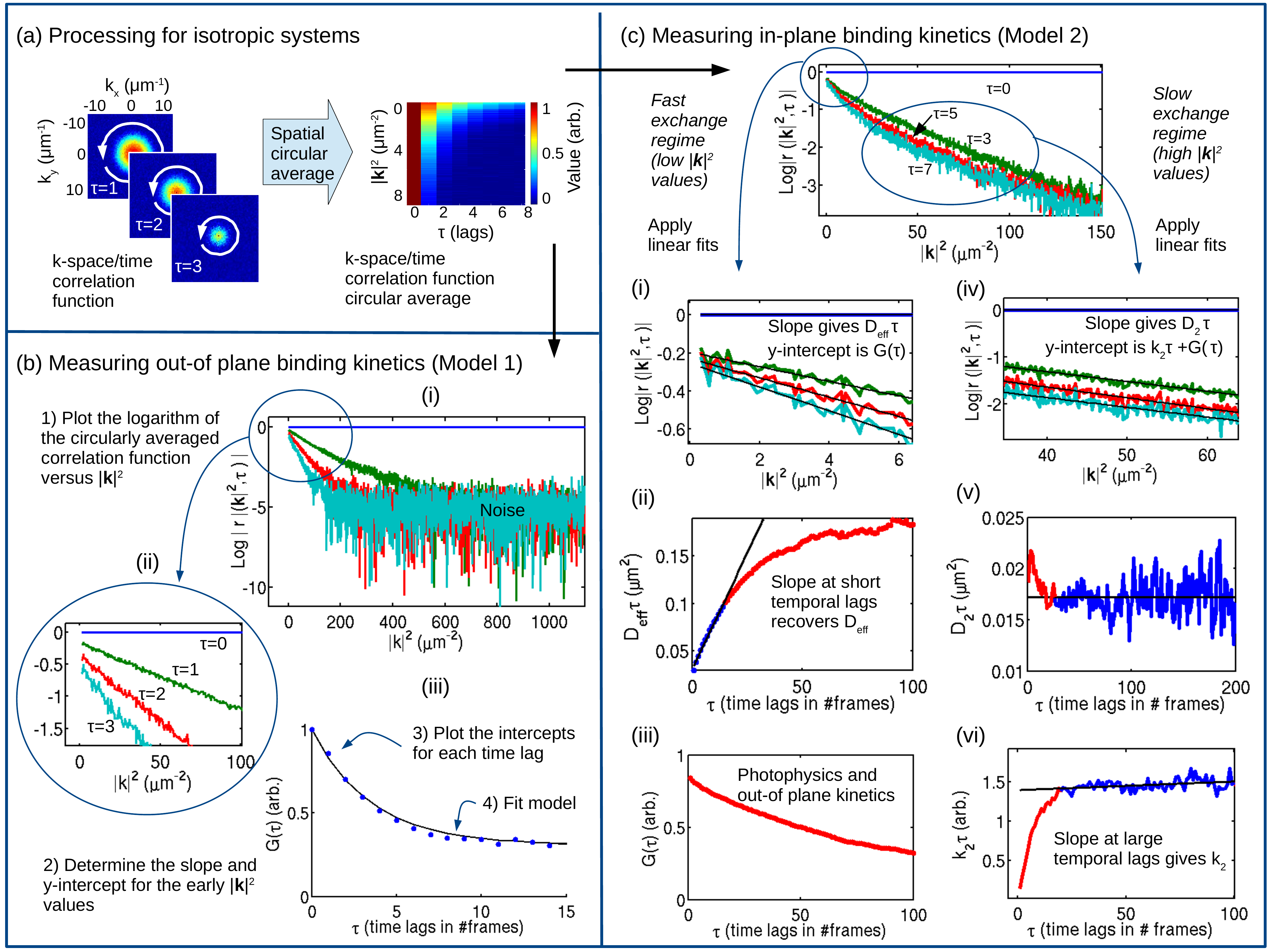}
\caption{Methods for extracting kinetic and dynamics parameters from the k-space/time correlation function. (a) Circular averaging of the kICS correlation function is performed for systems where diffusion is assumed isotropic; for each time lag of the correlation function, pixels equidistant from the central $|{\bf k}|^2=0$ point are averaged out. (b) Data processing steps to measure the kinetic parameters of ligand-receptor binding kinetics as in Fig~\ref{fig::kinetic_mechanisms}~a). Sample results are shown for a 2D ligand-receptor simulation. Simulation input parameter values: $k_d=0.20~s^{-1}$, $k_a=0.10~s^{-1}$, $D_c = 0.010~\mu m^2/s$; Extracted parameter values:  $k_d=0.20~s^{-1}$, $k_a=0.89~s^{-1}$, $D_c = 0.010~\mu m^2/s$. (c) Data processing steps used to measure the kinetic exchange parameters for a two-state diffusive model as in Fig~\ref{fig::kinetic_mechanisms}~b). Sample results correspond to experimental data of fluorescently labelled Cholera toxin bound to GM1 lipid domains on the basal 
membrane of a \emph{COS-7} cell as described in Section~\ref{sec::Experimental}. The logarithm of the circularly averaged correlation function is fit for both the fast and slow-exchange regime to extract the effective diffusion coefficient and the unbinding kinetic rate $k_2$.}
\label{fig::methodology}
\end{figure*}

\section{Experimental Analysis Methodology}

In this section we outline the steps followed to obtain kinetic rate parameters of the two model systems in Fig.~\ref{fig::kinetic_mechanisms} from the k-space/time image correlation function. The first step is to calculate the 2D kICS correlation function from the 3D k-space/time correlation function; this is done for isotropic systems by circularly averaging about the ${\bf k} = 0$ point, as shown in Fig.~\ref{fig::methodology}~a).

\subsection{Ligand-receptor binding}
For each time lag, $\tau$, the y-intercepts from a linear fit to the logarithm of the correlation function at low $|{\bf k}|^2$ give the photophysics term $G(\tau)$, as shown in Fig.~\ref{fig::methodology}~b.i). From the slopes of the correlation function, we can recover the diffusion coefficient for the ligand/receptor complex. We usually omit the first 5-15 $|{\bf k}|^2$ values (not shown in Fig.~\ref{fig::methodology}) for the fits. Low temporal sampling and issues due to windowing and the discrete Fourier transform make the values of the correlation function for the first few ${\bf |k|}^2$ more uncertain. To calculate the kinetic parameters $k_{a}$ and $k_{d}$ from $G(\tau)$ we use the binding model (Eq.~\ref{eq::gtau_binding}) and a non-linear least squares routine; a sample fit is shown in Fig.~\ref{fig::methodology}~b.ii) for a simulated image time-series. 

\subsection{Receptor-receptor docking}

For small values of $|{\bf k}|^2$, the correlation function is fit with the fast exchange model (Eq.~\ref{eq::fast_exchange}) for each time lag, $\tau$. The linear regime of the correlation function is more easily identified on a semi-logarithmic scale, and a linear fit to the logarithm of the correlation function at each time lag is calculated, as shown in Fig.~\ref{fig::methodology}~c.i). The slopes for each fit are given by $- D_{\text{eff}} \tau$ and the y-intercepts are given by $\log|C_1 G(\tau)|$. A fit to the slopes $D_{\text{eff}} \tau$ in the linear regime corresponding to low temporal lags, as shown in Fig.~\ref{fig::methodology}~c.ii), gives the value of $D_{\text{eff}}$.

For large values of $|{\bf k}|^2$, the correlation function is fit with the slow exchange model (Eq.~\ref{eq::slow_exchange}). Again, a linear fit to the logarithm of the correlation function is calculated at each $\tau$, as shown in Fig.~\ref{fig::methodology}~c.iv). The slopes are given by $-D_2 \tau$ and y-intercepts by $\log |C_2 G(\tau)| -k_2(\tau)$. The slopes from the slow exchange model can be used to calculate $D_2$. In Fig.~\ref{fig::methodology}~c.v), we find that the slope is $\approx 0$, and only the mean is plotted. For each time lag, subtracting the slow exchange y-intercepts from the fast exchange y-intercepts (Fig.~\ref{fig::methodology}~c.iii) gives |$\log |C_1/C_2| +k_2 \tau$. This step assumes that the photophysics terms are the same for the two states, though a more complex photophysics model could be used. Then, $k_2$ is calculated with a fit to the large temporal lags in the linear regime, as shown in Fig.~\ref{fig::methodology}~c.vi).

\section{Experimental Methods} \label{sec::Experimental}

\subsection{Ligand-receptor binding}

To test the ligand-receptor binding model represented in Fig.~\ref{fig::kinetic_mechanisms}~a), we use an in-vitro experimental system. The ligands of interest are human Double cortin (DCX)  motor proteins which we observe dynamically binding and unbinding from microtubules immobilized on an antibody-coated substrate. 

\subsubsection{Microtubule and Doublecortin preparation}
Human Doublecortin (DCX, accession number NP\_835365) tagged with EGFP at the C-terminus was purified using sequential Ni-NTA and Streptactin columns as described previously \cite{Bechstedt2012}.
Tubulin was purified from juvenile bovine brain homogenates as described previously \cite{ashford1998}. Labeling of cycled tubulin with TAMRA (Invitrogen) was performed as described \cite{hyman1991}; fluorescently-labeled tubulin was typically used at a labelling ratio of 1:4 labeled:unlabeled tubulin dimers. Tubulin was stored at -80~$^\circ$C degrees in BRB80 (BRB80 buffer: 80 mM PIPES pH 6.9 (KOH), 1~mM EGTA, 1~mM MgCl$_2$  filtered (0.22~µM), degassed, and stored at −20~$^\circ$C). Microtubule polymerization in the presence of GTP followed by stabilization with paclitaxel: A polymerization mixture was prepared with BRB80 + 32~$\mu$M tubulin + 1~mM GTP + 4~mM MgCl$_2$ + 5\% DMSO. The mixture was incubated on ice for 5~min, followed by incubation at 37~$^{\circ}$C for 30 min. The polymerized microtubules were diluted into pre-warmed BRB80 + 10~$\mu$M paclitaxel, centrifuged at maximum speed in a Beckman Airfuge, and resuspended in BRB80 + 10~$\mu$M paclitaxel.

\subsubsection{TIRF imaging of DCX and microtubules} 
The single-molecule assay for DCX (2.5~nM) was performed as described \cite{gell2010}: Microtubules were adhered to glass surfaces of the microscope chambers (blocked with 1\% Pluronic F-127) using anti-TAMRA antibody (1:200, Invitrogen). DCX was introduced into the chamber containing the surface-immobilized microtubules in imaging buffer (BRB80 + 10~$\mu$M paclitaxel + 0.1~mg/mL $\beta$-Mercaptoethanol + 0.1~mg/mL BSA + 1:100 dilution of antifade reagents (glucose, glucose oxidase, catalase). The sample chambers were constructed using custom-machined mounts diagrammed in Gell et al. (2010). In brief, microscope coverslips were silanized as described \cite{helenius2006}. A 22$\times$22~mm coverslip and an 18$\times$18~mm glass were separated by double-sided tape such that a narrow channel was created for the exchange of solution. 

For the imaging, a Zeiss Axiovert Z1 microscope chassis, 100x 1.45 NA Plan-apo-chromat objective lens, and the Zeiss TIRF III slider was used. Diode-pumped solid-state lasers (Cobolt Jive, Cobolt Calypso) were coupled to fiber-optic cables in free space and introduced into the Zeiss slider. Images were recorded using an Andor iXon+ DV-897 EMCCD camera using Metamorph. Frames were acquired with 0.1~s exposure times using continuous imaging.

\subsection{Receptor-receptor docking}

For the receptor-receptor docking model of Fig.~\ref{fig::kinetic_mechanisms}~b), we use an in-vivo experimental system. The receptor of interest is the multivalent Cholera toxin subunit B (CTxB) which may aggregate into small clusters and/or dock onto the actin cytoskeleton near the plasma membrane. 

\subsubsection{Cell culture}
Cell culture experiments were conducted with~\emph{COS-7} cells, a kidney fibroblast-like cell line derived from the~\emph{African green monkey}~\cite{cos7atcc}. Cells were cultured and passaged in medium according to standard procedures prescribed for this cell line~\cite{cos7atcc}. Briefly, cells were grown in glucose (0.45~$\%$ w/v), sodium-pyruvate (0.15~$\%$ w/v) and l-glutamine (4~mM) enriched Dulbecco Modified Eagle Medium (DMEM), supplemented by 10~$\%$ fetal bovine serum (FBS). Confluent cells were passaged (diluted) approximately every 3 days. Five days before imaging, cells were passaged onto fibronectin coated glass (glass N$^{\circ}$.1.5 of thickness 0.16-0.19~mm) MatTek (MatTek Corporation) dishes. DMEM was replaced by a nutrient reduced medium (Opti-Mem I, Invitrogen) one day prior to the imaging to induce cell starvation. 

\subsubsection{Cell labelling}
Ganglioside GM1 phospholipids were labelled by Cholera toxin subunit B (CTxB) conjugated to either Alexa-488 or Alexa 594 (Molecular Probes). The imaging medium was composed of Hank's Balanced Salt Solution (HBSS) supplemented by 10 mM HEPES buffer. Prior to imaging, cells were incubated for 10~min at 37~$^{\circ}$C in 1~mL of the imaging medium containing 2~$\mu$L of a diluted solution (0.1~mg/mL) of CTxB-Alexa-488. This step of the labelling procedure ensures that most of the GM1's are saturated with unlabelled Cholera toxin. Cells were then rinsed three times with the imaging medium and, subsequently, 2~$\mu$L of a CTxB-Alexa-594 diluted solution (0.1~mg/mL) were added. Imaging was carried out 10~min after addition of CTxB-Alexa-594 after the equilibrium between bound and unbound CTxB was established.

\subsubsection{Cell treatment with drugs} 
In order to achieve the de-polymerization of the actin cytoskeleton, cells were incubated in the presence of 1~$\mu$M of latrunculin B (5~min) and 5~$\mu$M Cytochalasin D (5 min), respectively. All of the imaging measurements were performed within 30~min of the cell treatment before any significant morphological changes in the cell could occur.

\subsubsection{TIRF microscopy of living cells} 
The imaging was carried out using a commercially available system: the TIRF III Research platform (Carl Zeiss, Germany) on an Axio Observer Z1 microscope. The imaging was performed with a 100x Alpha-Plan APO oil immersion objective lens with a NA of 1.46. The microscope stage was equipped with an enclosed heating module (37~$^{\circ}$C). Prior to each imaging session, the lasers were warmed up for at least an hour to stabilize the output intensity. Beam collimation and the critical angle calibrations were performed as described in the Zeiss user manual. The system was equipped with a CCD camera (Evolve S12 EMCCD) capable of imaging 512 by 512 pixel areas. The pixel size was 0.1~$\mu$m and the frame time was set to 46 or 94~ms with continuous imaging. Solid state lasers lines of 488 nm with 100 mW output power and 561 nm with power output of 40 mW, were used to excite Alexa 488 and 594, respectively. GFP and Rhodamine emission filters ensured selection of green and red signal, respectively. Data were acquired 
using the AxioVert software customized for this microscope. Each image series consisted of 256 by 256 pixel images, acquired over 2000-5000 frames. For each condition (control and drugs) 4-6 image series were acquired sequentially. Data series were stored in ``.zvi'' files and loaded into Matlab for further analysis.

\section{Results and Discussion}

\subsection{Testing the ligand-receptor binding model}

Microtubules are cytoskeletal filaments that are essential for cell structure and function, particularly during mitosis, cell migration, and neuronal development. Doublecortin (DCX) is a microtubule-associated protein (MAP) expressed during brain development. Mutations in the DCX gene cause defects in the development of the cerebral cortex, namely X-linked lissencephaly and double-cortex syndrome ~\cite{Gleeson1998}.

In recent single-molecule studies, the interaction times of DCX binding to microtubules has been studied. Bechstedt and Brouhard showed that the life-time of the bound state of a microtubule and DCX molecule can be modelled as a ligand bound to multivalent receptor system~\cite{Bechstedt2012}. Here, we confirm these results by analysing similar single-molecule experimental data. 

\begin{figure}[!ht] 
\centering
\includegraphics[scale=0.91]{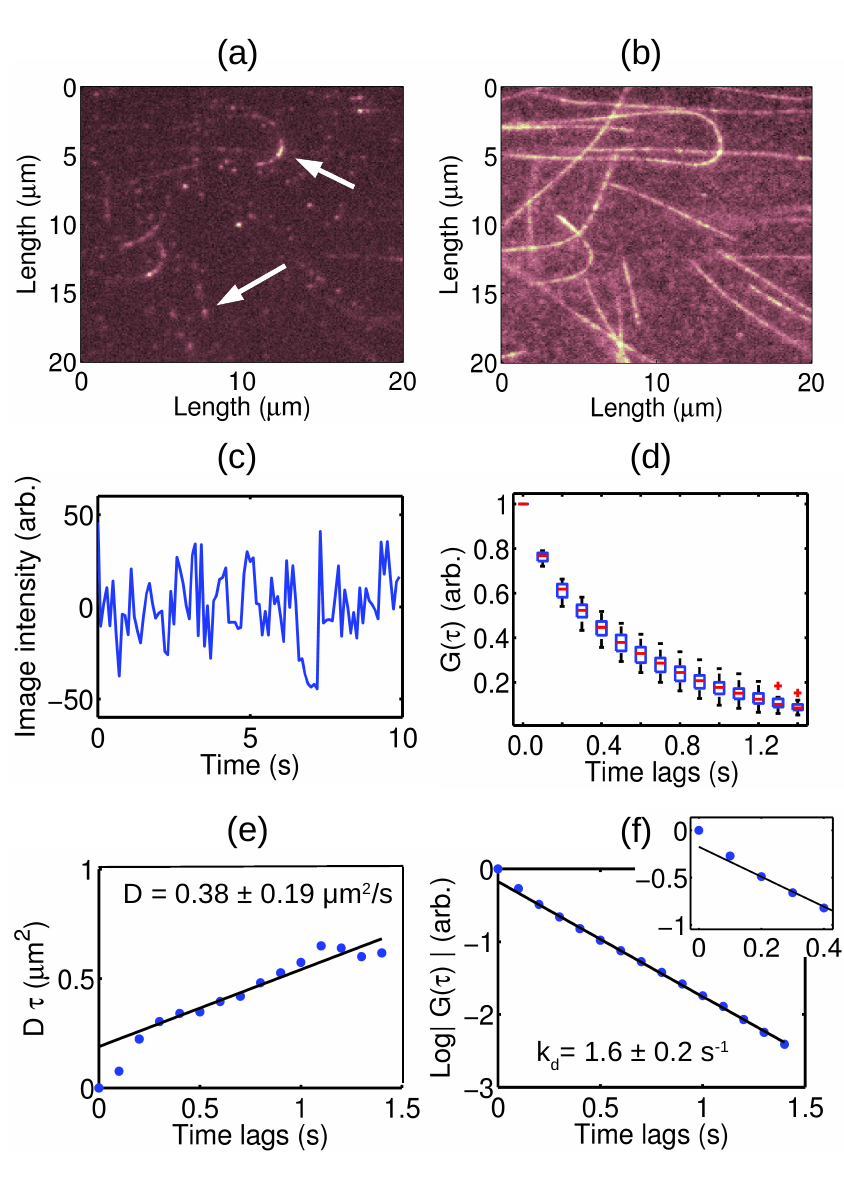}
\caption{Experimental results for a model system as shown in Fig.~\ref{fig::kinetic_mechanisms}~a): the kinetics and dynamics of Doublecortin binding to microtubules. (a) A sample region of interest of a single TIRF image (0.1~s exposure) of EGFP labelled DCX molecules (2.5nM) bound to microtubules on a sample substrate. The arrows are pointing to single bound molecules or small clusters of molecules. (b) Fluorescently labelled microtubules on the sample substrate imaged separetely in a different channel. An image series of the microtubules can serve as control measurements for microtubule movement, which may contribute to the apparent fluorescence fluctuations in $G(\tau)$. (c) Integrated image intensity versus time (minus the mean image intensity). This kind of measurement serves as an indicator of fluorescence photobleaching. (d) A box plot of the  experimentally measured kICS photophysics correlation function of DCX binding to microtubules. The solid (red) line in each box represents the median of a data set containing 17 image 
series, the box limits are the 25th and 75th percentiles and whiskers indicate the value of the most extreme points. (e) Plot of the mean-squared-displacement due to DCX diffusion averaged over all data series. (f) Plot of the logarithm of the mean photophysics correlation function with model fit as shown.}
\label{fig::Microtubule_Kinetics}
\end{figure}

We analysed in-vitro assays of DCX (2.5~nM) where microtubules were attached to a glass surface by antibodies. Fluorecent (EGFP labelled) DCX molecules were imaged in the green channel (Fig.~\ref{fig::Microtubule_Kinetics}~a) and were seen to dynamically bind and unbind from microtubules; fluorescent (TARMA labelled) tubulin was imaged in the red channel (Fig.~\ref{fig::Microtubule_Kinetics}~b). 

For our experiments, we analysed 17 datasets of 100 frames with 0.1~s exposure time for each frame; the total imaging time was 10~s per movie. Photobleaching effects were negligible for these data-sets and did not contribute to the overall photophysics correlations as seen in a sample intensity trace in Fig.~\ref{fig::Microtubule_Kinetics}~c) where the mean intensity has been removed. We performed our kICS analysis on data from both the green and red channels following the procedures outlined in Fig.~\ref{fig::kicsMethod}, Fig.~\ref{fig::methodology}~a) and Fig.~\ref{fig::methodology}~b) and we retrieved the experimental curve for the photophysics temporal correlation function, $G(\tau)$, as seen in Fig.~\ref{fig::Microtubule_Kinetics}~d). 

Using Eqs.~\ref{eq::gtau} and \ref{eq::gtau_binding} assuming $\Gamma_{fluor}(\tau)\approx 1$, we fit our kinetic binding model to obtain a kinetic dissociation rate of $k_d=1.6\pm0.2~s^{-1}$ (mean $\pm$ standard deviation); the association rate was too small to be accurately measured. Previously, Bechstedt and Brouhard had shown that DCX cooperatively interacts with microtubules such that higher concentrations of DCX will have higher average interaction lifetimes with their respective microtubules~\cite{Bechstedt2012}. In this previous study, the kinetic dissociation rate was measured at $k_d = 1.1~\pm~0.1~s^{-1}$ for a 10~nM DCX assay, and $k_d = 0.13~\pm~0.02~s^{-1}$ for a 2~$\mu$M concentration. Accounting for the cooperativity effects, our results, taken at 2.5~nM seem to be consistent with this previous finding.

From the kICS analysis of the dataset, we were also able to extract the diffusion coefficient of DCX. As seen from Fig.~\ref{fig::Microtubule_Kinetics}~e), we measure the diffusion coefficient to be $D=0.38\pm0.19~\mu m^2/s$ (mean $\pm$ standard deviation). To our knowledge, this is the first time such a measurement is reported for DCX; however, this diffusion coefficient is consistent with other single-particle tracking results on MAPs. A 2006 measurement on the kinesin-13 motor protein MCAK coincidentally also reports a diffusion coefficient of $0.38~\mu m^2/s$ similar to our results for DCX~\cite{helenius2006}. 

For two main reasons, in fitting the kinetic dissociation constant and the diffusion coefficient, we did not use the first two time-lags. First, kICS cannot distinguish between specific and non-specific binding of DCX onto the sample substrate/microtubule. DCX tends to aggregate into clusters in free solution and sometimes these appear in the imaging plane. However, these events decorrelate quickly (within a few frames) due to the significantly higher diffusion coefficient of the unbound molecules as seen by the deviation of the first two points from the linear fit to the logarithm of $G(\tau)$ in Fig.~\ref{fig::Microtubule_Kinetics}~f). Second, although the microtubules are bound to the glass coverslip by antibodies, some microtubules may have free ends which do not properly adhere. Due to thermal fluctuations, these microtubule segments move in and out of the imaging plane affecting the photophysics correlation function, $G(\tau)$; these also contribute to the small 
deviation seen in Fig.~\ref{fig::Microtubule_Kinetics}~f).

Separate kICS measurements carried out on the red channel can monitor how much microtubule movement will contribute to the correlation function. Likewise, control measurements of DCX imaged in the green channel without microtubules present will give an indication of the degree of non-specific binding. These can subsequently be used as references to correct $G(\tau)$. 

Finally, we must note some sources of systmatic error. As noted in the FCS literature, finite-length measurements will always introduce a small bias in the measured correlation function~\cite{Qian1990}~\cite{Saffarian2003}. The quantities of interest such as the kinetic binding rates and the diffusion coefficient will likely be overestimates of the true values since the finite-length correlation function appears to decorrelate faster. Furthermore, CCD camera noise will also similarly affect the measurement of $G(\tau)$; a systematic study of the effect has yet to be carried out, but we note that from simulations we can recover our input parameters typically to within 20\% error up to a signal to noise ratio of 3.

\subsection{Testing the receptor-receptor docking model}

\begin{figure}[!ht]
\centering
\includegraphics[scale=1.1]{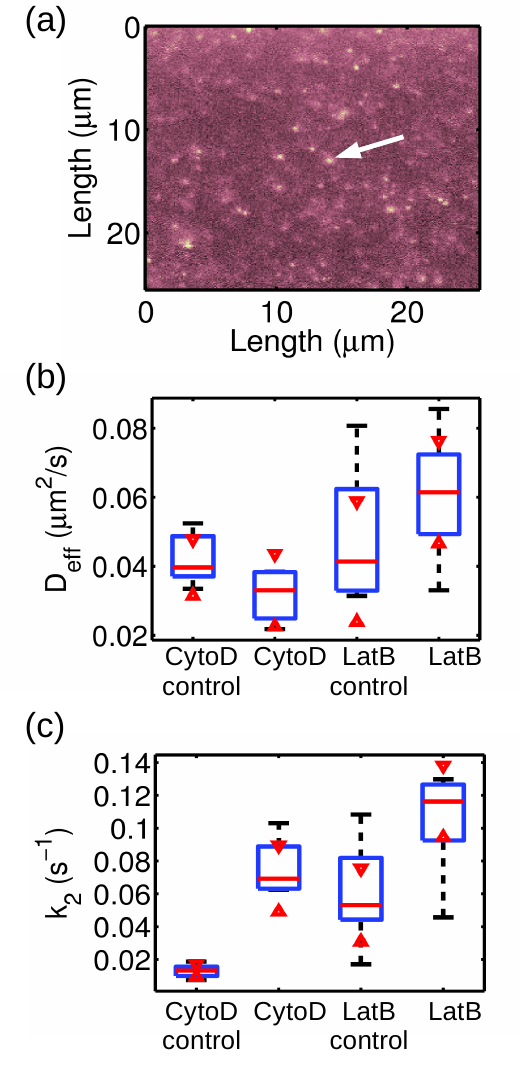}
\caption{Experimental results for a model system like that shown in Fig.~\ref{fig::kinetic_mechanisms}~b): the kinetics and transport dynamics of Cholera toxin bound to GM1 lipid domains in living cells. (a) A sample fluorescence image of Cholera toxin molecules bound to GM1 lipid domains. The arrow points to a slowly diffusing cluster of Cholera toxin molecules; these are likely to be vesicles transporting Cholera toxin into the cytoplasm. Fainter fluorescent spots are the freely diffusing Cholera toxin/GM1 complexes which appear to switch between two diffusing states. (b) The measured effective diffusion coefficient extracted using the method of Fig.~\ref{fig::methodology}~c). The (red) bar is the median of the dataset and the box limits represent the 25th and 75th percentiles while whiskers are the extreme points. The triangles are the confidence bounds for significance at $p<0.05$. (c) Measured kinetic rate constants of the Cholera toxin/GM1 complex representing the change from the more slowly to faster 
diffusing states. Actin inhibitors and disruption drug treatments reveal significant differences in the measured kinetic dissociation rates $k_2$.}
\label{fig::Cholera_Toxin_Kinetics}
\end{figure}

The B-subunit of Cholera Toxin (CTxB), belonging to the AB5 family of bacterial toxins, has the ability to bind to the cell membrane by cross linking glycolipids, specifically the glycolipid receptor GM1~\cite{Pina2005}. The binding of CTxB to its receptor is regulated through cholesterol dependent pathways ~\cite{Lingwood2011}, and the presence of the CTxB/GM1 complex in turn changes the membrane structure ~\cite{Hammond2005}~\cite{Bacia2005}.

It has been shown that the CTxB has a much slower diffusion compared to several other proteins and lipids ~\cite{Bacia2004}~ \cite{kenworthy2004}~ \cite{Goins1986}, though the question of how the cell regulates the dynamics of the CTx/GM1 complex has not been fully answered. There are several possible mechanisms which may regulate the dynamics of the complex; Day and Kenworthy recently confirmed, using confocal-fluorescence recovery after photobleaching (confocal-FRAP), that disrupting the actin cytoskeleton increases the mobility of CTxB on live cells relative to other lipids and proteins ~\cite{Day2012}. They reported a statistically significant increase in the diffusion coefficient on COS-7 cells, from $D = 0.21\pm0.1~\mu m^2/s$ to $0.35\pm0.18~\mu m^2/s$ after disruption of the actin cytoskeleton by Latrunculin A, suggesting that the actin cytoskeleton confines CTxB.

Here we use the CTxB/GM1 and the actin system to illustrate the kICS method as seen in Fig.~\ref{fig::Cholera_Toxin_Kinetics}~a). To investigate the effect of the actin cytoskeleton on the the complex, we disrupted actin with 1$\mu$M Latrunculin B and 5~$\mu$M Cytochalasin D, both of which act by blocking its polymerization ~\cite{wakatsuki2001}. We calculate the effective diffusion coefficient, $D_{\text{eff}}$, of the CTxGM1 complex, as well as the rate, $k_2$, at which the complex converts from a ‘docked state’ (possibly a glycolipid cluster) to a diffusing state. 

Our results for the diffusion coefficient are smaller than those reported previously using confocal-FRAP ~\cite{Bacia2004}~ \cite{kenworthy2004}~ \cite{Goins1986}~\cite{Day2012}. In contrast to the free diffusion coefficient reported by Day and Kenworthy $D = 0.21\pm0.1~\mu m^2/s$, we find an effective diffusion coefficient $D_{\text{eff}} = 0.046\pm0.015~\mu m^2/s$ (mean $\pm$ standard deviation) for control cells with no drug treatment. However, FRAP measures the diffusion of the mobile state at shorter length scales (i.e. on the size of the bleach spot which is $4.1~\mu m$ in diameter), and kICS measures the effective diffusion coefficient (which is an average diffusion of both mobile and docked states) at large length scales ($25.6~\mu m$ in length). As such, the measurements are not directly comparable. 

Upon drug treatments, we do not find a statistically significant change in our effective diffusion coefficients (as seen in Fig.~\ref{fig::Cholera_Toxin_Kinetics}~b) which suggests that on large spatial length scales a different (actin-independent) confinement mechanism may be more prominent; this may include confinement in ATP-dependent domains which have been shown to reduce to the mobility of the complex ~\cite{Day2012}.

The dissociation rate, $k_{2}$, was found to have significantly increased for both treatments (at $p<0.05$ for 6-8 cells using a Student's t-test) as seen in Fig.~\ref{fig::Cholera_Toxin_Kinetics}~c). The observed change in the kinetic rate is an increase of $k_{2}=0.06\pm 0.03~s^{-1}$ to $0.10\pm0.03~s^{-1}$ for Latrunculin B, and from $k_2=0.013\pm0.004~s^{-1}$ to  $0.08\pm0.02~ s^{-1}$ for Cytochalasin D. Since the disruptions increase the rate at which the CTxB/GM1 complex converts from a slowly diffusing state to a more quickly diffusing state, this suggests that the complex is confined by the actin cytoskeleton over short length scales. 

Finally, it is interesting to note that in image series with no noticeable photo-bleaching, we find a slowly decaying photophysics term $G(\tau)$. We believe that this decay may be due partly to vesicle dynamics and uptake of CTxB into the cell.

\section{Conclusion}

We introduced an extension of the theory of k-space image correlation spectroscopy that incorporated chemical binding and showed how to measure the kinetic rate constants and transport dynamics for two model experimental systems. This method offers an advantage over other techniques because in k-space, the kinetic binding rates may be determined independently of spatio-temporal parameters such as the diffusion coefficients and molecular flow velocities in the limit of large image sizes, or whenever molecules are confined such that lateral escape from the image region of interest is not allowed. Measurements of the average residency time of human Doublecortin bound to microtubules in an in-vitro system were consistent with single particle tracking data published previously. Furthermore, by perturbing the cytoskeleton with various drug treatments, we showed that our technique can measure significant differences in the kinetic binding rates of Cholera toxin/GM1 lipid complexes to the actin 
cytoskeleton. Our results suggest that the kICS technique can be applied to study different biochemical signalling pathways using both in-vivo and in-vitro systems.

\section{Acknowledgements}

We would like to thank Dr. David Ronis for his comments and critiques at the start of this project and Dr. Asmahan Abu-Arish for discussions involving the Cholera toxin experimental system. Furthermore, we thank Dr. Gergely Lukacs for offering the use of his TIRF microscope. H.B.B is grateful for funding by NSERC for a Canada Graduate Scholarship (CGS-M). G.J.B. and S.B. were supported by operating grants from the Canadian Institutes of Health Research (CIHR MOP-111265) and the Natural Sciences and Engineering Research Council (NSERC \#372593-09). P.W.W. acknowledges funding from the Natural Sciences and Engineering Research Council (Canada) Discovery Grant and the Canadian Institutes for Heath Research (CIHR).

\bibliographystyle{model1-num-names}
\bibliography{References.bib}

\begin{thebibliography}{45}
\expandafter\ifx\csname natexlab\endcsname\relax\def\natexlab#1{#1}\fi
\providecommand{\url}[1]{\texttt{#1}}
\providecommand{\href}[2]{#2}
\providecommand{\path}[1]{#1}
\providecommand{\DOIprefix}{doi:}
\providecommand{\ArXivprefix}{arXiv:}
\providecommand{\URLprefix}{URL: }
\providecommand{\Pubmedprefix}{pmid:}
\providecommand{\doi}[1]{\href{http://dx.doi.org/#1}{\path{#1}}}
\providecommand{\Pubmed}[1]{\href{pmid:#1}{\path{#1}}}
\providecommand{\bibinfo}[2]{#2}
\ifx\xfnm\relax \def\xfnm[#1]{\unskip,\space#1}\fi
\bibitem[{Feinerman et~al.(2008)Feinerman, Germain, and
  Altan-Bonnet}]{Feinerman2008}
\bibinfo{author}{O.~Feinerman}, \bibinfo{author}{R.~N. Germain},
  \bibinfo{author}{G.~Altan-Bonnet},
\newblock \bibinfo{title}{{Quantitative challenges in understanding ligand
  discrimination by $\alpha \beta$ T cells}},
\newblock \bibinfo{journal}{Molecular Immunology} \bibinfo{volume}{45}
  (\bibinfo{year}{2008}) \bibinfo{pages}{619--631}.
\bibitem[{Govern et~al.(2010)Govern, Paczosa, Chakraborty, and
  Huseby}]{Govern2010}
\bibinfo{author}{C.~C. Govern}, \bibinfo{author}{M.~K. Paczosa},
  \bibinfo{author}{A.~K. Chakraborty}, \bibinfo{author}{E.~S. Huseby},
\newblock \bibinfo{title}{{Fast on-rates allow short dwell time ligands to
  activate T cells}},
\newblock \bibinfo{journal}{PNAS} \bibinfo{volume}{107} (\bibinfo{year}{2010})
  \bibinfo{pages}{8724--8729}.
\bibitem[{Dushek et~al.(2009)Dushek, Das, and Coombs}]{Dushek2009}
\bibinfo{author}{O.~Dushek}, \bibinfo{author}{R.~Das},
  \bibinfo{author}{D.~Coombs},
\newblock \bibinfo{title}{{A role for rebinding in rapid and reliable T cell
  responses to antigen}},
\newblock \bibinfo{journal}{PLoS computational biology} \bibinfo{volume}{5}
  (\bibinfo{year}{2009}) \bibinfo{pages}{1--12}.
\bibitem[{Kholodenko et~al.(1999)Kholodenko, Demin, Moehren, and
  Hoek}]{Kholodenko1999}
\bibinfo{author}{B.~N. Kholodenko}, \bibinfo{author}{O.~V. Demin},
  \bibinfo{author}{G.~Moehren}, \bibinfo{author}{J.~B. Hoek},
\newblock \bibinfo{title}{{Quantification of short term signaling by the
  epidermal growth factor receptor}},
\newblock \bibinfo{journal}{The Journal of biological chemistry}
  \bibinfo{volume}{274} (\bibinfo{year}{1999}) \bibinfo{pages}{30169--30181}.
\bibitem[{de~Curtis(2001)}]{Curtis2001}
\bibinfo{author}{I.~de~Curtis},
\newblock \bibinfo{title}{{Cell migration: GAPs between membrane traffic and
  the cytoskeleton}},
\newblock \bibinfo{journal}{EMBO reports} \bibinfo{volume}{2}
  (\bibinfo{year}{2001}) \bibinfo{pages}{277--81}.
\bibitem[{Kaufmann et~al.(2012)Kaufmann, Strasser, and Jost}]{Kaufmann2012}
\bibinfo{author}{T.~Kaufmann}, \bibinfo{author}{A.~Strasser},
  \bibinfo{author}{P.~J. Jost},
\newblock \bibinfo{title}{{Fas death receptor signalling: roles of Bid and
  XIAP}},
\newblock \bibinfo{journal}{Cell death and differentiation}
  \bibinfo{volume}{19} (\bibinfo{year}{2012}) \bibinfo{pages}{42--50}.
\bibitem[{Elmore(2007)}]{Elmore2007}
\bibinfo{author}{S.~Elmore},
\newblock \bibinfo{title}{{Apoptosis: a review of programmed cell death}},
\newblock \bibinfo{journal}{Toxicologic pathology} \bibinfo{volume}{35}
  (\bibinfo{year}{2007}) \bibinfo{pages}{495--516}.
\bibitem[{Llopis et~al.(2012)Llopis, Sliusarenko, Heinritz, and
  Jacobs-Wagner}]{MonteroLlopis2012}
\bibinfo{author}{P.~M. Llopis}, \bibinfo{author}{O.~Sliusarenko},
  \bibinfo{author}{J.~Heinritz}, \bibinfo{author}{C.~Jacobs-Wagner},
\newblock \bibinfo{title}{{In vivo biochemistry in bacterial cells using FRAP:
  Insight into the translation cycle}},
\newblock \bibinfo{journal}{Biophysical Journal} \bibinfo{volume}{103}
  (\bibinfo{year}{2012}) \bibinfo{pages}{1848--1859}.
\bibitem[{Yasuda et~al.(2006)Yasuda, Harvey, Zhong, Sobczyk, van Aelst, and
  Svoboda}]{Yasuda2006}
\bibinfo{author}{R.~Yasuda}, \bibinfo{author}{C.~D. Harvey},
  \bibinfo{author}{H.~Zhong}, \bibinfo{author}{A.~Sobczyk},
  \bibinfo{author}{L.~van Aelst}, \bibinfo{author}{K.~Svoboda},
\newblock \bibinfo{title}{{Supersensitive Ras activation in dendrites and
  spines revealed by two-photon fluorescence lifetime imaging.}},
\newblock \bibinfo{journal}{Nature neuroscience} \bibinfo{volume}{9}
  (\bibinfo{year}{2006}) \bibinfo{pages}{283--291}.
\bibitem[{Michelman-Ribeiro et~al.(2009)Michelman-Ribeiro, Mazza, Rosales,
  Stasevich, Boukari, Rishi, Vinson, Knutson, and
  McNally}]{Michelman-Ribeiro2009}
\bibinfo{author}{A.~Michelman-Ribeiro}, \bibinfo{author}{D.~Mazza},
  \bibinfo{author}{T.~Rosales}, \bibinfo{author}{T.~J. Stasevich},
  \bibinfo{author}{H.~Boukari}, \bibinfo{author}{V.~Rishi},
  \bibinfo{author}{C.~Vinson}, \bibinfo{author}{J.~R. Knutson},
  \bibinfo{author}{J.~G. McNally},
\newblock \bibinfo{title}{{Direct measurement of association and dissociation
  rates of DNA binding in live cells by fluorescence correlation
  spectroscopy}},
\newblock \bibinfo{journal}{Biophysical journal} \bibinfo{volume}{97}
  (\bibinfo{year}{2009}) \bibinfo{pages}{337--346}.
\bibitem[{Stasevich et~al.(2010)Stasevich, Mueller, Michelman-Ribeiro, Rosales,
  Knutson, and McNally}]{Stasevich2010}
\bibinfo{author}{T.~J. Stasevich}, \bibinfo{author}{F.~Mueller},
  \bibinfo{author}{A.~Michelman-Ribeiro}, \bibinfo{author}{T.~Rosales},
  \bibinfo{author}{J.~R. Knutson}, \bibinfo{author}{J.~G. McNally},
\newblock \bibinfo{title}{{Cross-validating FRAP and FCS to quantify the impact
  of photobleaching on in vivo binding estimates.}} \bibinfo{volume}{99}
  (\bibinfo{year}{2010}) \bibinfo{pages}{3093--101}.
\bibitem[{Elson and Magde(1974)}]{Elson1974}
\bibinfo{author}{E.~L. Elson}, \bibinfo{author}{D.~Magde},
\newblock \bibinfo{title}{{Fluorescence correlation spectroscopy. I. Conceptual
  basis and theory}},
\newblock \bibinfo{journal}{Biopolymers} \bibinfo{volume}{13}
  (\bibinfo{year}{1974}) \bibinfo{pages}{1--27}.
\bibitem[{Magde et~al.(1972)Magde, Elson, and Webb}]{Magde1972}
\bibinfo{author}{D.~Magde}, \bibinfo{author}{E.~Elson}, \bibinfo{author}{W.~W.
  Webb},
\newblock \bibinfo{title}{{Thermodynamic Fluctuations in a Reacing System -
  Measurement by FCS}},
\newblock \bibinfo{journal}{Physical Review Letters} \bibinfo{volume}{29}
  (\bibinfo{year}{1972}) \bibinfo{pages}{705--708}.
\bibitem[{Thompson et~al.(1981)Thompson, Burghardt, and Axelrod}]{Thompson1981}
\bibinfo{author}{N.~L. Thompson}, \bibinfo{author}{T.~P. Burghardt},
  \bibinfo{author}{D.~Axelrod},
\newblock \bibinfo{title}{{Measuring surface dynamics of biomolecules by total
  internal reflection fluorescence with photobleaching recovery or correlation
  spectroscopy}},
\newblock \bibinfo{journal}{Biophysical journal} \bibinfo{volume}{33}
  (\bibinfo{year}{1981}) \bibinfo{pages}{435--454}.
\bibitem[{Thompson and Axelrod(1983)}]{Thompson1983}
\bibinfo{author}{N.~L. Thompson}, \bibinfo{author}{D.~Axelrod},
\newblock \bibinfo{title}{{Immunoglobulin surface-binding kinetics studied by
  total internal reflection with fluorescence correlation spectroscopy.}},
\newblock \bibinfo{journal}{Biophysical journal} \bibinfo{volume}{43}
  (\bibinfo{year}{1983}) \bibinfo{pages}{103--114}.
\bibitem[{Starr and Thompson(2001)}]{Starr2001}
\bibinfo{author}{T.~E. Starr}, \bibinfo{author}{N.~L. Thompson},
\newblock \bibinfo{title}{{Total internal reflection with fluorescence
  correlation spectroscopy: combined surface reaction and solution diffusion}},
\newblock \bibinfo{journal}{Biophysical journal} \bibinfo{volume}{80}
  (\bibinfo{year}{2001}) \bibinfo{pages}{1575--84}.
\bibitem[{Thompson et~al.(2011)Thompson, Navaratnarajah, and
  Wang}]{Thompson2011}
\bibinfo{author}{N.~L. Thompson}, \bibinfo{author}{P.~Navaratnarajah},
  \bibinfo{author}{X.~Wang},
\newblock \bibinfo{title}{{Measuring surface binding thermodynamics and
  kinetics by using total internal reflection with fluorescence correlation
  spectroscopy: practical considerations}},
\newblock \bibinfo{journal}{The journal of physical chemistry. B}
  \bibinfo{volume}{115} (\bibinfo{year}{2011}) \bibinfo{pages}{120--131}.
\bibitem[{Kolin and Wiseman(2007)}]{Kolin2007}
\bibinfo{author}{D.~L. Kolin}, \bibinfo{author}{P.~W. Wiseman},
\newblock \bibinfo{title}{{Advances in image correlation spectroscopy:
  measuring number densities, aggregation states, and dynamics of fluorescently
  labeled macromolecules in cells.}},
\newblock \bibinfo{journal}{Cell biochemistry and biophysics}
  \bibinfo{volume}{49} (\bibinfo{year}{2007}) \bibinfo{pages}{141--64}.
\bibitem[{Kannan et~al.(2007)Kannan, Guo, Sudhaharan, Ahmed, Maruyama, and
  Wohland}]{Kannan2007}
\bibinfo{author}{B.~Kannan}, \bibinfo{author}{L.~Guo},
  \bibinfo{author}{T.~Sudhaharan}, \bibinfo{author}{S.~Ahmed},
  \bibinfo{author}{I.~Maruyama}, \bibinfo{author}{T.~Wohland},
\newblock \bibinfo{title}{{Spatially Resolved Total Internal Reflection
  Fluorescence Correlation Microscopy Using an Electron Multiplying
  Charge-Coupled Device Camera}},
\newblock \bibinfo{journal}{Analytical chemistry} \bibinfo{volume}{79}
  (\bibinfo{year}{2007}) \bibinfo{pages}{4463--4470}.
\bibitem[{Sankaran et~al.(2009)Sankaran, Manna, Guo, Kraut, and
  Wohland}]{Sankaran2009}
\bibinfo{author}{J.~Sankaran}, \bibinfo{author}{M.~Manna},
  \bibinfo{author}{L.~Guo}, \bibinfo{author}{R.~Kraut},
  \bibinfo{author}{T.~Wohland},
\newblock \bibinfo{title}{{Diffusion, transport, and cell membrane organization
  investigated by imaging fluorescence cross-correlation spectroscopy.}},
\newblock \bibinfo{journal}{Biophysical journal} \bibinfo{volume}{97}
  (\bibinfo{year}{2009}) \bibinfo{pages}{2630--2639}.
\bibitem[{Hebert et~al.(2005)Hebert, Costantino, and Wiseman}]{Hebert2005}
\bibinfo{author}{B.~Hebert}, \bibinfo{author}{S.~Costantino},
  \bibinfo{author}{P.~W. Wiseman},
\newblock \bibinfo{title}{{Spatiotemporal image correlation spectroscopy
  (STICS) theory, verification, and application to protein velocity mapping in
  living CHO cells.}},
\newblock \bibinfo{journal}{Biophysical journal} \bibinfo{volume}{88}
  (\bibinfo{year}{2005}) \bibinfo{pages}{3601--3014}.
\bibitem[{Toplak et~al.(2012)Toplak, Pandzic, Chen, Vicente-Manzanares,
  Horwitz, and Wiseman}]{Toplak2012}
\bibinfo{author}{T.~Toplak}, \bibinfo{author}{E.~Pandzic},
  \bibinfo{author}{L.~Chen}, \bibinfo{author}{M.~Vicente-Manzanares},
  \bibinfo{author}{A.~Horwitz}, \bibinfo{author}{P.~Wiseman},
\newblock \bibinfo{title}{{STICCS Reveals Matrix-Dependent Adhesion Slipping
  and Gripping in Migrating Cells}},
\newblock \bibinfo{journal}{Biophysical Journal} \bibinfo{volume}{103}
  (\bibinfo{year}{2012}) \bibinfo{pages}{1672--1682}.
\bibitem[{Tanaka and Papoian(2010)}]{Tanaka2010}
\bibinfo{author}{N.~Tanaka}, \bibinfo{author}{G.~A. Papoian},
\newblock \bibinfo{title}{{Reverse-engineering of biochemical reaction networks
  from spatio-temporal correlations of fluorescence fluctuations.}},
\newblock \bibinfo{journal}{Journal of theoretical biology}
  \bibinfo{volume}{264} (\bibinfo{year}{2010}) \bibinfo{pages}{490--500}.
\bibitem[{Kolin et~al.(2006)Kolin, Ronis, and Wiseman}]{Kolin2006}
\bibinfo{author}{D.~L. Kolin}, \bibinfo{author}{D.~Ronis},
  \bibinfo{author}{P.~W. Wiseman},
\newblock \bibinfo{title}{{k-Space image correlation spectroscopy: a method for
  accurate transport measurements independent of fluorophore photophysics.}},
\newblock \bibinfo{journal}{Biophysical journal} \bibinfo{volume}{91}
  (\bibinfo{year}{2006}) \bibinfo{pages}{3061--3075}.
\bibitem[{Hamming(1989)}]{Hamming}
\bibinfo{author}{R.~Hamming}, \bibinfo{title}{Digital Filters},
  \bibinfo{edition}{3rd} ed., \bibinfo{publisher}{Prentice-Hall},
  \bibinfo{address}{Englewood Cliffs, New Jersey}, \bibinfo{year}{1989}.
\bibitem[{Lauffenburger and Linderman(1993)}]{Lauffenburger}
\bibinfo{author}{D.~A. Lauffenburger}, \bibinfo{author}{J.~L. Linderman},
  \bibinfo{title}{Receptors: Models for Binding, Trafficking, and Signalling},
  \bibinfo{publisher}{Oxford University Press}, \bibinfo{address}{New York, New
  York}, \bibinfo{year}{1993}.
\bibitem[{Berne and Pecora(2000)}]{BernePecora}
\bibinfo{author}{B.~J. Berne}, \bibinfo{author}{R.~Pecora},
  \bibinfo{title}{Dynamic Light Scattering: With Applications to Chemistry,
  Biology, and Physics}, \bibinfo{publisher}{Dover Publications},
  \bibinfo{address}{Mineola, New York}, \bibinfo{year}{2000}.
\bibitem[{Bechstedt and Brouhard(2012)}]{Bechstedt2012}
\bibinfo{author}{S.~Bechstedt}, \bibinfo{author}{G.~J. Brouhard},
\newblock \bibinfo{title}{Doublecortin recognizes the 13-protofilament
  microtubule cooperatively and tracks microtubule ends},
\newblock \bibinfo{journal}{Dev Cell} \bibinfo{volume}{23}
  (\bibinfo{year}{2012}) \bibinfo{pages}{181--192}.
\bibitem[{Ashford et~al.(1998)Ashford, Andersen, and Hyman}]{ashford1998}
\bibinfo{author}{A.~Ashford}, \bibinfo{author}{S.~Andersen},
  \bibinfo{author}{A.~A. Hyman},
\newblock \bibinfo{title}{{Preparation of Tubulin from Bovine Brain}},
\newblock in: \bibinfo{editor}{J.~Celis} (Ed.), \bibinfo{booktitle}{Cell
  Biology, A Laboratory Handbook}, volume~\bibinfo{volume}{2},
  \bibinfo{publisher}{Academic Press}, \bibinfo{address}{New York},
  \bibinfo{year}{1998}, pp. \bibinfo{pages}{205--212}.
\bibitem[{Hyman et~al.(1991)Hyman, Drechsel, Kellogg, Salser, Sawin, Steffen,
  Wordeman, and Mitchison}]{hyman1991}
\bibinfo{author}{A.~Hyman}, \bibinfo{author}{D.~Drechsel},
  \bibinfo{author}{D.~Kellogg}, \bibinfo{author}{S.~Salser},
  \bibinfo{author}{K.~Sawin}, \bibinfo{author}{P.~Steffen},
  \bibinfo{author}{L.~Wordeman}, \bibinfo{author}{T.~Mitchison},
\newblock \bibinfo{title}{{Preparation of modified tubulins}},
\newblock \bibinfo{journal}{Methods Enzymol} \bibinfo{volume}{196}
  (\bibinfo{year}{1991}) \bibinfo{pages}{478--85}.
\bibitem[{Gell et~al.(2010)Gell, Bormuth, Brouhard, Cohen, Diez, Friel,
  Helenius, Nitzsche, Petzold, Ribbe, Schaffer, Stear, Trushko, Varga, Widlund,
  Zanic, and Howard}]{gell2010}
\bibinfo{author}{C.~Gell}, \bibinfo{author}{V.~Bormuth}, \bibinfo{author}{G.~J.
  Brouhard}, \bibinfo{author}{D.~N. Cohen}, \bibinfo{author}{S.~Diez},
  \bibinfo{author}{C.~T. Friel}, \bibinfo{author}{J.~Helenius},
  \bibinfo{author}{B.~Nitzsche}, \bibinfo{author}{H.~Petzold},
  \bibinfo{author}{J.~Ribbe}, \bibinfo{author}{E.~Schaffer},
  \bibinfo{author}{J.~H. Stear}, \bibinfo{author}{A.~Trushko},
  \bibinfo{author}{V.~Varga}, \bibinfo{author}{P.~O. Widlund},
  \bibinfo{author}{M.~Zanic}, \bibinfo{author}{J.~Howard},
\newblock \bibinfo{title}{{Microtubule dynamics reconstituted in vitro and
  imaged by single-molecule fluorescence microscopy}},
\newblock \bibinfo{journal}{Methods Cell Biol} \bibinfo{volume}{95}
  (\bibinfo{year}{2010}) \bibinfo{pages}{221--45}.
\bibitem[{Helenius et~al.(2006)Helenius, Brouhard, Kalaidzidis, Diez, and
  Howard}]{helenius2006}
\bibinfo{author}{J.~Helenius}, \bibinfo{author}{G.~Brouhard},
  \bibinfo{author}{Y.~Kalaidzidis}, \bibinfo{author}{S.~Diez},
  \bibinfo{author}{J.~Howard},
\newblock \bibinfo{title}{{The depolymerizing kinesin MCAK uses lattice
  diffusion to rapidly target microtubule ends}},
\newblock \bibinfo{journal}{Nature} \bibinfo{volume}{441}
  (\bibinfo{year}{2006}) \bibinfo{pages}{115--9}.
\bibitem[{ATCC(2013)}]{cos7atcc}
\bibinfo{author}{ATCC}, \bibinfo{title}{Cell culture methods and general
  information for {C}os-7 ({ATCC CRL-1651})}, \bibinfo{year}{2013}. \URLprefix
  \url{http://www.atcc.org/Products/All/CRL-1651.aspx}.
\bibitem[{Gleeson et~al.(1998)Gleeson, Allen, Fox, Lamperti, Berkovic,
  Scheffer, Cooper, Dobyns, Minnerath, Ross, and Walsh}]{Gleeson1998}
\bibinfo{author}{J.~Gleeson}, \bibinfo{author}{K.~Allen},
  \bibinfo{author}{J.~Fox}, \bibinfo{author}{E.~Lamperti},
  \bibinfo{author}{S.~Berkovic}, \bibinfo{author}{I.~Scheffer},
  \bibinfo{author}{E.~Cooper}, \bibinfo{author}{W.~Dobyns},
  \bibinfo{author}{S.~Minnerath}, \bibinfo{author}{M.~Ross},
  \bibinfo{author}{C.~Walsh},
\newblock \bibinfo{title}{Doublecortin, a brain-specific gene mutated in human
  x-linked lissencephaly and double cortex syndrome, encodes a putative
  signaling protein},
\newblock \bibinfo{journal}{Cell} \bibinfo{volume}{92} (\bibinfo{year}{1998})
  \bibinfo{pages}{63--72}.
\bibitem[{Qian(1990)}]{Qian1990}
\bibinfo{author}{H.~Qian},
\newblock \bibinfo{title}{On the statistics of fluorescence correlation
  spectroscopy},
\newblock \bibinfo{journal}{Biophysical Chemistry} \bibinfo{volume}{38}
  (\bibinfo{year}{1990}) \bibinfo{pages}{49--57}.
\bibitem[{Saffarian and Elson(2003)}]{Saffarian2003}
\bibinfo{author}{S.~Saffarian}, \bibinfo{author}{E.~L. Elson},
\newblock \bibinfo{title}{Statistical analysis of fluorescence correlation
  spectroscopy: The standard deviation and bias},
\newblock \bibinfo{journal}{Biophysical Journal} \bibinfo{volume}{84}
  (\bibinfo{year}{2003}) \bibinfo{pages}{2030–2042}.
\bibitem[{Pina and Johannes(2005)}]{Pina2005}
\bibinfo{author}{D.~G. Pina}, \bibinfo{author}{L.~Johannes},
\newblock \bibinfo{title}{Cholera and shiga toxin b-subunits: thermodynamic and
  structural considerations for function and biomedical applications},
\newblock \bibinfo{journal}{Toxicon} \bibinfo{volume}{45}
  (\bibinfo{year}{2005}) \bibinfo{pages}{389 -- 393}.
\bibitem[{Lingwood et~al.(2011)Lingwood, Binnington, Rog, Vattulainen, Grzybek,
  Coskun, Lingwood, and Simons}]{Lingwood2011}
\bibinfo{author}{D.~Lingwood}, \bibinfo{author}{B.~Binnington},
  \bibinfo{author}{T.~Rog}, \bibinfo{author}{I.~Vattulainen},
  \bibinfo{author}{M.~Grzybek}, \bibinfo{author}{U.~Coskun},
  \bibinfo{author}{C.~A. Lingwood}, \bibinfo{author}{K.~Simons},
\newblock \bibinfo{title}{{{C}holesterol modulates glycolipid conformation and
  receptor activity}},
\newblock \bibinfo{journal}{Nat. Chem. Biol.} \bibinfo{volume}{7}
  (\bibinfo{year}{2011}) \bibinfo{pages}{260--262}.
\bibitem[{Hammond et~al.(2005)Hammond, Heberle, Baumgart, Holowka, Baird, and
  Feigenson}]{Hammond2005}
\bibinfo{author}{A.~T. Hammond}, \bibinfo{author}{F.~A. Heberle},
  \bibinfo{author}{T.~Baumgart}, \bibinfo{author}{D.~Holowka},
  \bibinfo{author}{B.~Baird}, \bibinfo{author}{G.~W. Feigenson},
\newblock \bibinfo{title}{{{C}rosslinking a lipid raft component triggers
  liquid ordered-liquid disordered phase separation in model plasma
  membranes}},
\newblock \bibinfo{journal}{Proc. Natl. Acad. Sci. U.S.A.}
  \bibinfo{volume}{102} (\bibinfo{year}{2005}) \bibinfo{pages}{6320--6325}.
\bibitem[{Bacia et~al.(2005)Bacia, Schwille, and Kurzchalia}]{Bacia2005}
\bibinfo{author}{K.~Bacia}, \bibinfo{author}{P.~Schwille},
  \bibinfo{author}{T.~Kurzchalia},
\newblock \bibinfo{title}{{{S}terol structure determines the separation of
  phases and the curvature of the liquid-ordered phase in model membranes}},
\newblock \bibinfo{journal}{Proc. Natl. Acad. Sci. U.S.A.}
  \bibinfo{volume}{102} (\bibinfo{year}{2005}) \bibinfo{pages}{3272--3277}.
\bibitem[{Bacia et~al.(2004)Bacia, Scherfeld, Kahya, and Schwille}]{Bacia2004}
\bibinfo{author}{K.~Bacia}, \bibinfo{author}{D.~Scherfeld},
  \bibinfo{author}{N.~Kahya}, \bibinfo{author}{P.~Schwille},
\newblock \bibinfo{title}{{{F}luorescence correlation spectroscopy relates
  rafts in model and native membranes}},
\newblock \bibinfo{journal}{Biophys. J.} \bibinfo{volume}{87}
  (\bibinfo{year}{2004}) \bibinfo{pages}{1034--1043}.
\bibitem[{Kenworthy et~al.(2004)Kenworthy, Nichols, Remmert, Hendrix, Kumar,
  Zimmerberg, and Lippincott-Schwartz}]{kenworthy2004}
\bibinfo{author}{A.~K. Kenworthy}, \bibinfo{author}{B.~J. Nichols},
  \bibinfo{author}{C.~L. Remmert}, \bibinfo{author}{G.~M. Hendrix},
  \bibinfo{author}{M.~Kumar}, \bibinfo{author}{J.~Zimmerberg},
  \bibinfo{author}{J.~Lippincott-Schwartz},
\newblock \bibinfo{title}{{{D}ynamics of putative raft-associated proteins at
  the cell surface}},
\newblock \bibinfo{journal}{J. Cell Biol.} \bibinfo{volume}{165}
  (\bibinfo{year}{2004}) \bibinfo{pages}{735--746}.
\bibitem[{Goins et~al.(1986)Goins, Masserini, Barisas, and Freire}]{Goins1986}
\bibinfo{author}{B.~Goins}, \bibinfo{author}{M.~Masserini},
  \bibinfo{author}{B.~G. Barisas}, \bibinfo{author}{E.~Freire},
\newblock \bibinfo{title}{{{L}ateral diffusion of ganglioside {G}{M}1 in
  phospholipid bilayer membranes}},
\newblock \bibinfo{journal}{Biophys. J.} \bibinfo{volume}{49}
  (\bibinfo{year}{1986}) \bibinfo{pages}{849--856}.
\bibitem[{Day and Kenworthy(2012)}]{Day2012}
\bibinfo{author}{C.~A. Day}, \bibinfo{author}{A.~K. Kenworthy},
\newblock \bibinfo{title}{Mechanisms underlying the confined diffusion of
  cholera toxin b-subunit in intact cell membranes},
\newblock \bibinfo{journal}{PLoS ONE} \bibinfo{volume}{7}
  (\bibinfo{year}{2012}) \bibinfo{pages}{e34923}.
\bibitem[{Wakatsuki et~al.(2001)Wakatsuki, Schwab, Thompson, and
  Elson}]{wakatsuki2001}
\bibinfo{author}{T.~Wakatsuki}, \bibinfo{author}{B.~Schwab},
  \bibinfo{author}{N.~C. Thompson}, \bibinfo{author}{E.~L. Elson},
\newblock \bibinfo{title}{{{E}ffects of cytochalasin {D} and latrunculin {B} on
  mechanical properties of cells}},
\newblock \bibinfo{journal}{J. Cell. Sci.} \bibinfo{volume}{114}
  (\bibinfo{year}{2001}) \bibinfo{pages}{1025--1036}.

\end{thebibliography}

\appendix

\section{Kinetics out of excitation plane} \label{sec::One_component_appendix}
Consider a simple two-state model, where a system is composed of free ligands in solution (fluorescent), free receptors on the cellular membrane (non-fluorescent), and bound ligand-receptor complexes. The reaction mechanism for the simple two state model is given by:

\begin{equation}
[L]+[R] \underset{k_d}{\overset{k_a}{\rightleftarrows}} [C]
\end{equation}
\noindent where $[L]$, $[R]$, $[C]$ are the ligand , receptor, and ligand-receptor complex concentrations respectively. 

The rate equations are given simply by:
\begin{equation} \label{eq::matrix_rate_equation}
\frac{d}{dt}\begin{pmatrix}
R \\ C
\end{pmatrix} =
\begin{pmatrix}
-k_a L & k_d \\
k_a L & -k_d
\end{pmatrix}
 \begin{pmatrix}
R \\ C
\end{pmatrix}
\end{equation}
\noindent where we have assumed that $[L]$ is approximately constant. Solving the matrix equation is fairly straight-forward and it gives the solution:

\begin{equation}
C(t) = \frac{k_a[L]}{k_d+k_a[L]} + \frac{k_d}{k_d+k_a[L]} e^{-(k_d+k_a[L])t} 
\end{equation}

\noindent for initial conditions $[C]=1$, $[R]=0$. The binding correlation function is proportional to: 

\begin{equation}
\langle \psi_{bind}(0) \psi_{bind}(\tau)\rangle \propto \langle C(0)C(\tau) \rangle.
\end{equation}

If we normalize $C(t)$, as was already done, such that $C(0) = 1$, then we give a probabilistic interpretation to concentration fluctuations. This way, $\langle \psi_{bind}(0) \psi_{bind}(\tau)\rangle = \langle C(0)C(\tau) \rangle$ and we recover Eq.~\ref{eq::gtau_binding}. For simplicity, we absorb $[L]$ into the definition of $k_a$.

\section{Interaction kinetics in-plane } \label{sec::Two_component_appendix}

Similar to the procedure for the kICS derivation for two non-interacting populations followed by Kolin et al.~\cite{Kolin2006}, we can write an image series $i(\textbf{r},t)$ with two fluorescent populations as:

\begin{equation}
i(\textbf{r},t) = q_1 I(\textbf{r}) * \rho_1(\textbf{r},t) + q_2 I(\textbf{r}) * \rho_2(\textbf{r},t),
\end{equation}

\noindent where $\rho_{1}(\textbf{r},t)$ and $\rho_{2}(\textbf{r},t)$ are the number densities of the fluorescent particles of each population at point $\textbf{r}$ at time t, $q_{1}$ and $q_{2}$  are their respective quantum yields, $I(\textbf{r})$ is the instrument point spread function, and the asterisk, $*$, represents a convolution integral. We model the number density of population $m$ as $\rho_m(\textbf{r},t)$ as:

 \begin{equation}
\rho_m(\textbf{r},t) = \sum_{i=1}^{N_m} \Theta _{m,i}(t)\delta(\textbf{r}-\textbf{r}_i(t)),
\end{equation}
\noindent where $\delta$ is the Dirac delta-function and the sum is over all $N_m$ particles of the $m$'th population and $\Theta _{m,i}(t)$ represents the photophysical state of the $i$'th particle, where $\Theta (t) = 1$ for a visible, fluorescing particle and $\Theta (t)=0$ for a photobleached, or non-fluorescing particle at time $t$. The 2D spatial Fourier transform of the image series is then:
\begin{equation}
\tilde{i}(\textbf{k},t) = \tilde{I}(\textbf{k}) \left[q_1\sum_{i=1}^{N_1} \Theta_{1,i}(t) e^{i\textbf{k} \cdot \textbf{r}_i(t)}+ q_2\sum_{j=1}^{N_2} \Theta_{2,j}(t)e^{i\textbf{k} \cdot \textbf{r}_j(t)}
\right],
\end{equation}
\noindent where the tilde represents the spatial Fourier transform. 

The k-space/time correlation function $r(\textbf{k};\tau,t)$ in this case becomes,

\begin{equation} \label{2poprkt}
\begin{split}
& r(\textbf{k};\tau,t) =  |\tilde{I}(\textbf{k})|^2  ~~\times \\ &
    \Big\langle \left(q_1\sum_{i=1}^{N_1} \Theta_{1,i}(t) e^{i\textbf{k} \cdot \textbf{r}_i(t)} + q_2\sum_{j=1}^{N_2} \Theta_{2,j}(t) e^{i\textbf{k} \cdot \textbf{r}_j(t)} \right)  ~~\times \\ &
     \left( q_1\sum_{i=1}^{N_1} \Theta_{1,i}(t+\tau) e^{-i\textbf{k} \cdot \textbf{r}_i(t+\tau)} + q_2\sum_{j=1}^{N_2} \Theta_{2,j}(t+\tau) e^{-i\textbf{k} \cdot \textbf{r}_j(t+\tau)}\right) \Big\rangle
\end{split}
\end{equation}

\noindent where the angular brackets indicate an ensemble average over all possible particle configurations. We now make a simplifying assumption: we consider dilute-enough concentrations of particles such that they only correlate with themselves. Allowing for self-correlations arising from particles converting between diffusing states, the expansion of Eq.~(\ref{2poprkt}) gives:

\begin{equation}\label{eq::full_expansion}
\begin{split}
r(\textbf{k};\tau,t) &= |\tilde{I}(\textbf{k})|^2
    [ q_1^2\langle\Theta_1(t)\Theta_1(t+\tau)\rangle\langle\tilde{c}_1(\textbf{k},t)\tilde{c}_1(\textbf{k},t+\tau)\rangle\\
    &+ q_1q_2\langle\Theta_1(t)\Theta_2(t+\tau)\rangle\langle\tilde{c}_1(\textbf{k},t)\tilde{c}_2(\textbf{k},t+\tau)\rangle\\
    &+ q_2q_1\langle\Theta_2(t)\Theta_1(t+\tau)\rangle\langle\tilde{c}_2(\textbf{k},t)\tilde{c}_1(\textbf{k},t+\tau)\rangle\\
    &+ q_2^2\langle\Theta_2(t)\Theta_2(t+\tau)\rangle\langle\tilde{c}_2(\textbf{k},t)\tilde{c}_2(\textbf{k},t+\tau)\rangle]
\end{split}
\end{equation}

 \noindent where $\tilde{c}_m(\textbf{k},t)$ is the concentration of the population $m$ in Fourier space and it is proportional to the $e^{i {\bf k} \cdot r(t)}$ terms and the number of particles in the given population. The photophysics correlations, $\langle\Theta_m(t)\Theta_n(t+\tau)\rangle$, have been separated from the concentration correlations, $\langle\tilde{c}_m(\textbf{k},t)\tilde{c}_n(\textbf{k},t+\tau)\rangle$, by assuming the statistical independence of photophysics from particle dynamics.

 The first and fourth of the four terms are the self-correlation functions for populations 1 and 2, while the second and third terms are the cross correlation terms between the two populations. Eq.~(\ref{eq::full_expansion}) is re-written more compactly as:

 \begin{equation} \label{eq::compact_expansion}
 r(\textbf{k};\tau,t) = |\tilde{I}(\textbf{k})|^2 \sum_{m=1}^2 \sum_{n=1}^2 \alpha_{mn}\textbf{F}_{mn}(\textbf{k},\tau)
 \end{equation}

 \noindent where $\alpha_{mn} = q_mq_n\langle\Theta_m(t)\Theta_n(t+\tau)\rangle$ are the photophysics terms and $\textbf{F}_{mn}=\langle\tilde{c}_m(\textbf{k},t)\tilde{c}_n(\textbf{k},t+\tau)\rangle$ are the particle correlation terms.
$\textbf{F}_{mn}$ can be found for our system of interest by considering the two component reaction diffusion system given by the kinetic rate equations:

\begin{align}
\label{eq::Dc1}
\frac{\partial c_1(\textbf{r},t))}{\partial t} &= D_1 \nabla^2 c_1(\textbf{r},t) - k_1 c_1(\textbf{r},t)  \\ \nonumber & \hspace{10pt}+ k_2 c_2(\textbf{r},t)  \label{eq::Dc2} \\
\frac{\partial c_2(\textbf{r},t)}{\partial t} &= D_2 \nabla^2 c_2(\textbf{r},t) + k_1 c_1(\textbf{r},t) \\ \nonumber & \hspace{10pt}- k_2 c_2(\textbf{r},t).
\end{align}

We assume throughout this derivation that $D_1 > D_2$, and these quantities are sufficiently different to be distinguishable. Following the derivation by Berne and Pecora ~\cite{BernePecora}, the solution to this system is found by taking the Fourier-Laplace transforms of Eqs.~(\ref{eq::Dc1}) and~(\ref{eq::Dc2}) which gives:

\begin{eqnarray}
\hat{{F}}_{11}(k,s) &=&  \bar{c}_1 \frac{[s+\gamma_2]}{\Delta(s)} \\
\hat{{F}}_{21}(k,s) &=&  \bar{c}_2 \frac{k_2}{\Delta(s)} \\
\hat{{F}}_{12}(k,s) &=&  \bar{c}_1 \frac{k_1}{\Delta(s)} \\
\hat{{F}}_{22}(k,s) &=&  \bar{c}_2 \frac{[s+\gamma_1]}{\Delta(s)}
\end{eqnarray}

\noindent where the hat (~\textasciicircum~) represents the Fourier-Laplace transform, the overbar on $ \bar{c}_m $ represents the equilibrium concentration of the $m$'th particle population, and $\Delta(s)$ is the determinant of the system of equations~(\ref{eq::Dc1}) and ~(\ref{eq::Dc2}) given by

\begin{equation}
\begin{split}
\Delta(s) &= (s + \gamma_1 )(s+\gamma_2) - k_1k_2 \\
       &= s^2 + s\gamma_1+s\gamma_2 + \gamma_1\gamma_2 -k_1k_2 \\
       &= s^2 + s |\textbf{k}|^2(D_1+D_2) +s(k_1 +k_2) \\& \hspace{20pt} + |\textbf{k}|^4D_1 D_2 + |\textbf{k}|^2 (D_1k_2 + D_2k_1)
\end{split}
\end{equation}

\noindent and $\gamma_1$ and $\gamma_2$, the transport coefficients given by

\begin{align}
\gamma_1 = D_1|{\textbf k}|^2 + k_1, \\
\gamma_2 = D_2|{\textbf k}|^2 + k_2.
\end{align}

Substituting the inverse Laplace transforms of the decoupled equations into Eq.~(\ref{eq::compact_expansion})  gives the correlation function in Fourier space:

\begin{equation}
r({\textbf k}, \tau) = B_+ e^{s_+ \tau} + B_- e^{s_- \tau} \label{twoexpon}
\end{equation}

\noindent  where $s_{\pm}$, the roots of $\Delta(s)$, are given by

\begin{equation}
s_{\pm} = -\frac{1}{2}(\gamma_1 + \gamma_2) \pm \frac{1}{2}[(\gamma_1 - \gamma_2)^2 + 4k_1k_2]^{\frac{1}{2}}
\end{equation}

\noindent while the coefficients are given by
\begin{align}
B_\pm & =  |\tilde{I}(\textbf{k})|^2  \frac{\alpha_{11} \bar{c}_1 (s_\pm + \gamma_2) + \alpha_{12} \bar{c}_1 k_1}{s_+ - s_-} \\ \nonumber & + |\tilde{I}(\textbf{k})|^2  \frac{\alpha_{21} \bar{c}_2 k_2 + \alpha_{22} \bar{c}_2 (s_\pm + \gamma_1)}{s_+ - s_-}
\end{align}

\noindent The exponents $s_{\pm}$ are dependent on the spatial Fourier frequency through the transport coefficients $\gamma_1$ and $\gamma_2$, while the coeffiecients are also time dependent through the photophysics terms $\alpha_{ij}$.

In the fast exchange regime, where $(k_1 + k_2) >> D_1 |{\bf{k}}|^2$, a first order perturbative expansion of $\Delta(s)$ (in the small quantity proportional to $|{\bf k}|^2D_1$) has the form $s \approx s^{(0)} + s^{(1)}$, where the bracketed superscript indicates the order of the expansion correction. This expansion gives the zeroth and first order corrections:

\begin{align}
0 &= [s^{(0)}]^2 + (k_1 + k_2)s^{(0)},\\
0 &= 2s^{(0)}s^{(1)} + (k_1+k_2)s^{(1)} +|\textbf{k}|^2(D_1+D_2)s^{(0)} \\ \nonumber & \hspace{10pt} +k_2|\textbf{k}|^2D_1+k_1|\textbf{k}|^2D_2
\end{align}

The solutions to these equations are:

\begin{align}
s_+ &= -|\textbf{k}|^2D_{\text{eff}} \\
s_- &= -(k_1+k_2) - |\textbf{k}|^2 (D_1 + D_2 - D_\text{eff})
\end{align}

\noindent and
\begin{align}
B_+ &= |\tilde{I}(\textbf{k})|^2  (k_1+k_2)^{-1} [ \alpha_{11}\bar{c}_1k_2 + \alpha_{12} \bar{c}_1k_1 \\ \nonumber & \hspace{15pt}+ \alpha_{21} \bar{c}_2k_2 +\alpha_{22} \bar{c}_2k_1 ] \\
B_- &= |\tilde{I}(\textbf{k})|^2 (k_1+k_2)^{-1} [- \alpha_{11} \bar{c}_1k_1 + \alpha_{12} \bar{c}_1k_1  \\ \nonumber& \hspace{15pt}+ \alpha_{21} \bar{c}_2k_2 -\alpha_{22} \bar{c}_2k_2 ]
\end{align}

\noindent where $D_{\text{eff}} = \frac{D_1k_2 + D_2k_1}{k_1 + k_2}$ defines the effective diffusion coefficient.

To remove the PSF dependence from $B_+$ and $B_-$, we normalize the correlation function, Eq.~(\ref{twoexpon}), by its zeroth time-lag. If we make the assumption that the photophysics of both diffusing states is about the same, then $G(\tau) = \alpha_{11} = \alpha_{12} = \alpha_{21} = \alpha_{22}$, the coefficient $B_-$ becomes zero, and we are left with the fast exchange correlation function:

\begin{equation}
 r_{fast}(|{\bf{k}}|^2,\tau) = C_1 ~ G(\tau)  \exp(-|{\bf k}|^2D_{\mathrm{eff}}\tau)\\
\end{equation}
\noindent where $C_1$ is just a constant. 

In the slow exchange regime, where $(k_1 + k_2) << D_1 |{\bf{k}}|^2$, a similar first order perturbative expansion of $\Delta(s)$ (in the small quantities $k_1$ and $k_2$) gives:

\begin{align}
0 &= [s^{(0)}]^2 + |\textbf{k}|^2(D_1+D_2)s^{(0)} \\
0 &= 2s^{(0)}s^{(1)} + (k_1+k_2)s^{(0)} +|\textbf{k}|^2(D_1+D_2)s^{(1)} \\ \nonumber &\hspace{10pt}+ k_2|\textbf{k}|^2D_1+k_1|\textbf{k}|^2D_2
\end{align}

\noindent which has the solution:

\begin{align}
s_+ &= -|\textbf{k}|^2D_{1} - k_1 \\
s_- &= -|\textbf{k}|^2D_{2} -k_2
\end{align}

\noindent and
\begin{align}
B_+ &= |\tilde{I}(\textbf{k})|^2 \alpha_{11}\bar{c}_1 \\
B_- &= |\tilde{I}(\textbf{k})|^2 \alpha_{22}\bar{c}_2
\end{align}

Again, after normalizing by the zeroth time-lag, the full correlation function in this regime then takes the form:

\begin{equation}
\begin{split}
 r(|{\bf{k}}|^2,\tau) &= C_2 \alpha_{11} \exp(-|{\bf k}|^2D_1\tau-k_1\tau) \\ &+ C_2 \alpha_{22} \exp(-|{\bf k}|^2D_2\tau-k_2\tau)
\end{split}
\end{equation}

At large $|{\bf{k}}|^2$ and $\tau$, assuming $D_1 >> D_2$ the first term decays quickly compared to the second term, leaving a simpler form of the normalized slow exchange correlation function:

\begin{equation}
 r_{slow}(|{\bf{k}}|^2,\tau) = C_2 G(\tau) \exp(-|{\bf k}|^2D_2\tau-k_2\tau)\\
\end{equation}

\noindent where $G(\tau) = \alpha_{22}$ and $C_2$ is just a constant.

\end{document}